\documentclass[aps,pre,twocolumn,showpacs,superscriptaddress]{revtex4}
\usepackage{epsfig}
\usepackage{times}
\begin{document}

\title{Evolution of emotions on networks leads to the evolution of cooperation in social dilemmas}

\author{Attila Szolnoki}
\affiliation{Institute of Technical Physics and Materials Science, Research Centre for Natural Sciences, Hungarian Academy of Sciences, P.O. Box 49, H-1525 Budapest, Hungary}

\author{Neng-Gang Xie}
\affiliation{Department of Mechanical Engineering, Anhui University of Technology, Maanshan City 243002, China}

\author{Ye Ye}
\affiliation{Department of Mechanical Engineering, Anhui University of Technology, Maanshan City 243002, China}

\author{Matja{\v z} Perc}
\affiliation{Faculty of Natural Sciences and Mathematics, University of Maribor, Koro{\v s}ka cesta 160, SI-2000 Maribor, Slovenia}

\begin{abstract}
We show that the resolution of social dilemmas on random graphs and scale-free networks is facilitated by imitating not the strategy of better performing players but rather their emotions. We assume sympathy and envy as the two emotions that determine the strategy of each player by any given interaction, and we define them as probabilities to cooperate with players having a lower and higher payoff, respectively. Starting with a population where all possible combinations of the two emotions are available, the evolutionary process leads to a spontaneous fixation to a single emotional profile that is eventually adopted by all players. However, this emotional profile depends not only on the payoffs but also on the heterogeneity of the interaction network. Homogeneous networks, such as lattices and regular random graphs, lead to fixations that are characterized by high sympathy and high envy, while heterogeneous networks lead to low or modest sympathy but also low envy. Our results thus suggest that public emotions and the propensity to cooperate at large depend, and are in fact determined by the properties of the interaction network.
\end{abstract}

\maketitle

\section{Introduction}
Evolutionary games \cite{hofbauer_98} have recently received ample attention in the physics community, as it became obvious that methods of statistical physics can be used successfully to study also interactions that are more complex than just those between particles \cite{liggett_85}. Broadly classified as statistical physics of social dynamics \cite{castellano_rmp09}, these studies aim to elevate our understanding of collective phenomena in society on a level that is akin to the understanding we have about interacting particle systems. Within the theoretical framework of evolutionary games, the evolution of cooperation \cite{axelrod_84} is probably the most interesting collective phenomenon to study. Several evolutionary games constitute so-called social dilemmas \cite{macy_pnas02}, the most prominent of which is the prisoner's dilemma game, and in which understanding the evolution of cooperation still a grand challenge. Regardless of game particularities, a social dilemma implies that the collective wellbeing is at odds with individual success. An individual is therefore tempted to act so as to maximize her own profit, but at the same time neglecting negative consequences this has for the society as a whole. A frequently quoted consequence of such selfish actions is the ``tragedy of the commons'' \cite{hardin_g_s68}. While cooperation is regarded as the strategy leading away from the threatening social decline, it is puzzling why individuals would choose to sacrifice some fraction of personal benefits for the wellbeing of society.

According to Nowak \cite{nowak_s06}, five rules promote the evolution of cooperation. These are kin selection, direct and indirect reciprocity, network reciprocity, and group selection. Recent reviews \cite{szabo_pr07, schuster_jbp08, roca_plr09, perc_bs10} clearly attest to the fact that physics-inspired research has helped refine many of these concepts. In particular evolutionary games on networks, spurred on by the seminal discovery of spatial reciprocity \cite{nowak_n92b}, and subsequently by the discovery that scale-free networks strongly facilitate the evolution of cooperation \cite{santos_prl05, santos_pnas06}, are still receiving ample attention to this day \cite{ohtsuki_n06, gomez-gardenes_prl07, ohtsuki_prl07, fu_pre09, pusch_pre08, traulsen_pnas09, van-segbroeck_prl09, lee_s_prl11, poncela_pre11, buesser_pre12, yang_hx_epl12, chen_yz_pre12, pinheiro_njp12, allen_jtb12, kuperman_pre12, allen_jtb12b,li_q_njp12, tanimoto_pre12,fu_jsp12,portillo_pre12}. One of the most recent contributions to the subject concerns the assignment of cognitive skills to individuals that engage in evolutionary games on networks \cite{szolnoki_epl11, brede_epl11, vukov_njp12,brede_acs12, szolnoki_pre12}. The earliest forerunners to these advances can be considered strategies such as ``tit-for-tat'' \cite{nowak_n92a} and Pavlov \cite{wedekind_pnas96}, many of which were proposed already during the seminal experiments performed by Axelrod \cite{axelrod_s81}, and which assume individuals have cognitive skills that exceed those granted to them in the framework of classical game theory. It has recently been shown, for example, that incipient cognition solves several open question related to network reciprocity and that cognitive strategies are particularly fit to take advantage of the ability of heterogeneous networks to promote the evolution of cooperation \cite{vukov_njp12}.

Here we build on our previous work \cite{szolnoki_epl11}, where we have presented the idea that not strategies but rather emotions could be the subject of imitation during the evolutionary process. It is worth noting that the transmissive nature of positive and negative emotional states was already observed in \cite{hill_prsb10}, where it was concluded that humans really do adjust their emotions depending on their contacts in a social network. Moreover, the connection between intuition and willingness to cooperate was also tested in human experiments \cite{rand_n12}. It therefore is of interest to determine how the topology of the interaction network affects the spreading of emotions, which may in turn determine the level of cooperation. In the context of games on lattices, we have shown that imitating emotions such as goodwill and envy from the more successful players reinstalls imitation as a tour de force for resolving social dilemmas, even for games where the Nash equilibrium is a mixed phase. We have also argued that envy is an important inhibitor of cooperative behavior. We now revisit the snowdrift, stag-hunt and the prisoner's dilemma game on random graphs and scale-free networks, with the aim of determining the role of interaction heterogeneity within this framework. We focus on sympathy and envy as the two key emotions determining the emotional profile of each player, and we define them simply as the probability to cooperate with less and more successful opponents, respectively. Strategies thus become link-specific rather than player-specific, whereby the level of cooperation in the population can be determined by the average number of times players choose to cooperate. Interestingly, in agreement with a recent experiment, we find that network reciprocity plays a negligible role \cite{gracia-lazaro_srep12}. The outcome on regular random graphs is the same as reported previously for the square lattice, leading to the conclusion that the ability of cooperators to aggregate into spatially compact clusters is irrelevant. Only when degree heterogeneity is introduced to interaction networks, we find that the evolution of emotional profiles changes. As we will show, homogeneous networks lead to fixations that are characterized by high sympathy and high envy, while heterogeneous networks lead to low or modest sympathy and low envy. Network heterogeneity thus alleviates a key impediment to higher levels of cooperation on lattices and regular networks, namely envy, and by doing so opens the possibility to much more cooperative states even under extremely adverse conditions. From a different point of view, it can be argued that some topological features of interaction networks in fact determine the emotional profiles of players, and they do so in such a way that cooperation is the most frequently chosen strategy.

The remainder of this paper is organized as follows. First, we describe the mathematical model, in particular the protocol for the imitation of emotional profiles as well as the definition of social dilemmas on networks. Next we present the main results, whereas lastly we summarize and discuss their implications.

\section{Mathematical model}

The traditional setup of an evolutionary game assumes $N$ players occupying vertices of an interaction network. Moreover, each player $x$ having a pure strategy, cooperates $(s_x=C)$ or defects $(s_x=D)$ with all neighbors independently of their strategy and payoff. Here, instead of pure strategies, we introduce an emotional profile $(\alpha_x, \beta_x) \in [0.1]$ to each player, which characterizes the willingness to cooperate towards a neighbor in dependence on the other player's success that is quantified by the payoff value. More precisely, if the corresponding payoff values are $p_x$ and $p_y$ for players $x$ and $y$, respectively, then $\alpha_x$ determines the probability that player $x$ will cooperate with player $y$ in case of $p_x > p_y$. Conversely, when $p_x < p_y$, the parameter $\beta_x$ is the probability that player $x$ will cooperate with player $y$. In the rare case of equality ($p_x = p_y$), the corresponding probability is the average of $\alpha_x$ and $\beta_x$.

In this way the $(\alpha_x, \beta_x)$ pair thus determines how a given player $x$ will behave when facing less or a more successful opponent $y$. As described in the Introduction, the pair determines each player's sympathy and envy. If $\alpha_x = 1$ we say the player is completely sympathetic. Alternative interpretations such as goodwill and charity are also viable, given that player $x$ will always cooperate with less successful opponents. Similarly, if $\beta=1$ we say the player is not envious. Despite the fact that the opponents are more successful, she will always cooperate with them. As by $\alpha_x$, alternatives such as servility or proneness to brownnose or ``butter up'' appear fit as well. It is important to note that a player $x$ may simultaneously cooperate and defect towards neighbors $y$ and ${y\prime}$ if their payoffs are very different. Furthermore, player $x$ may adopt different strategies even if $p_y \approx p_{y\prime}$ due to the probabilistic nature of an emotional profile.

When two players engage in a round of an evolutionary game, we assume that mutual cooperation yields the reward $R$, mutual defection leads to punishment $P$, and the mixed choice gives the cooperator the sucker's payoff $S$ and the defector the temptation $T$. Within this traditional setup we have the prisoner's dilemma (PD) game if $T>R>P>S$, the snowdrift game (SG) if $T>R>S>P$, and the stag-hunt (SH) game if $R>T>P>S$, thus covering all three major social dilemma types where players can choose between cooperation and defection. Following common practice \cite{szabo_pr07}, we set $R = 1$ and $P=0$, thus leaving the remaining two payoffs to occupy $-1 \leq S \leq 1$ and $0 \leq T \leq 2$, as depicted schematically in Fig.~\ref{RRG}(a).

To begin, each player $x$ is assigned a random $(\alpha_x, \beta_x)$ pair and a payoff from the reachable $[kS,kT]$ interval, where $k$ denotes the average degree of players.
Subsequently, every payoff value is updated by considering the proper neighborhoods of a player and the actual emotional parameters. Importantly, after the accumulation of new payoffs, player $y$ cannot imitate a pure strategy from player $x$ but only its emotional profile, i.e., the $\alpha_x$ and/or $\beta_x$ value. Imitation is decided so that a randomly selected player $x$ first acquires its payoff $p_x$ by playing the game with all its $k_x$ neighbors, as defined by the interaction network. Note that $k_x$ is thus the degree of player $x$. Next, one randomly chosen partner of $x$, denoted by $y$, also acquires its payoff $p_y$ by playing the game with all its $k_y$ neighbors. Player $y$ then attempts to imitate the emotional profile of players $x$ with the probability $q=1/\{1+\exp[(p_y-p_x)/K]\}$, where $K$ determines the level of uncertainty by strategy adoptions \cite{szabo_pr07}. The latter can be attributed to errors in judgment due to mistakes and external influences that affect the evaluation of the opponent. Without loss of generality we set $K=0.5$, implying that better performing players are readily imitated, but it is not impossible to adopt the strategy of a player performing worse. Importantly, since the emotional profile consist of two parameters, two random numbers are drawn to enable independent imitation of $\alpha_x$ and $\beta_x$. This is vital to avoid potential artificial propagations of freak (extremely successful) $(\alpha_x, \beta_x)$ pairs. Technically, $100 \times 100$ $(\alpha_x, \beta_x)$ pairs were available at the start of the evolutionary process. Finally, after each imitation the payoff of player $y$ is updated using its new emotional profile, whereby each full Monte Carlo step involves all players having a chance to adopt the emotional profile from one of their neighbors once on average.

Prior to presenting the result of this model, it is important to note that there will almost always be a fixation of $(\alpha_x, \beta_x)$ pairs, i.e., irrespective of $T$ and $S$ only a single pair will eventually spread across the whole population. Once fixation occurs the evolutionary process stops. The characteristic probability of encountering cooperative behavior in the population, which is equivalent to the stationary fraction of cooperators $f_C$ in the traditional version of the game, can then be determined by means of averaging over the final states that emerge from different initial conditions. Exceptions to single $(\alpha, \beta)$ pair fixations are likely to occur for strongly heterogeneous networks, where more than one $(\alpha_x, \beta_x)$ pair can survive around strong hubs. This effect is more pronounced in the harmony game (HG) quadrant, but becomes negligible in the prisoner's dilemma parametrization of the game. In case more than a single $(\alpha_x, \beta_x)$ pair does survive, we present in what follows the average over several independent realizations. For the Monte Carlo simulations, we have used $N=5000-40000$ players and up to $10^7$ full steps, and we have averaged over $100-500$ independent runs.

\section{Results}

\begin{figure}
\centerline{\epsfig{file=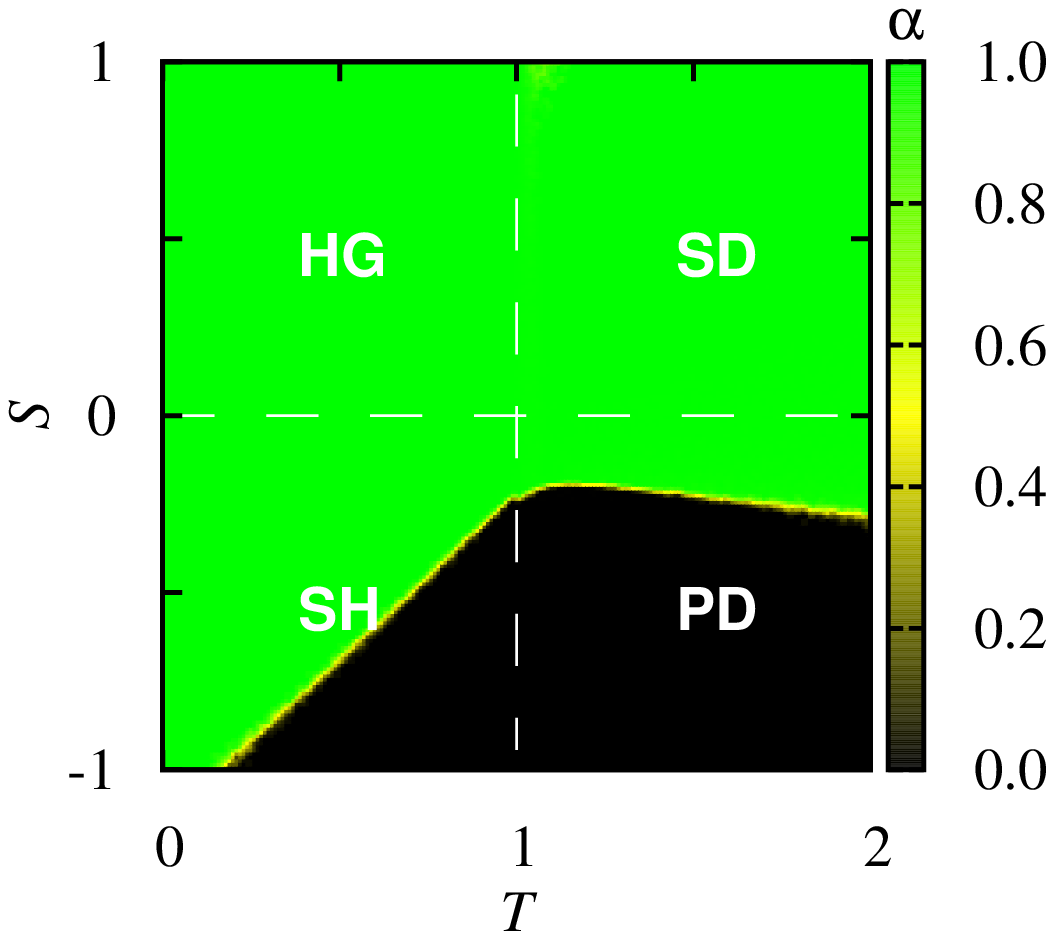,width=4.3cm}\epsfig{file=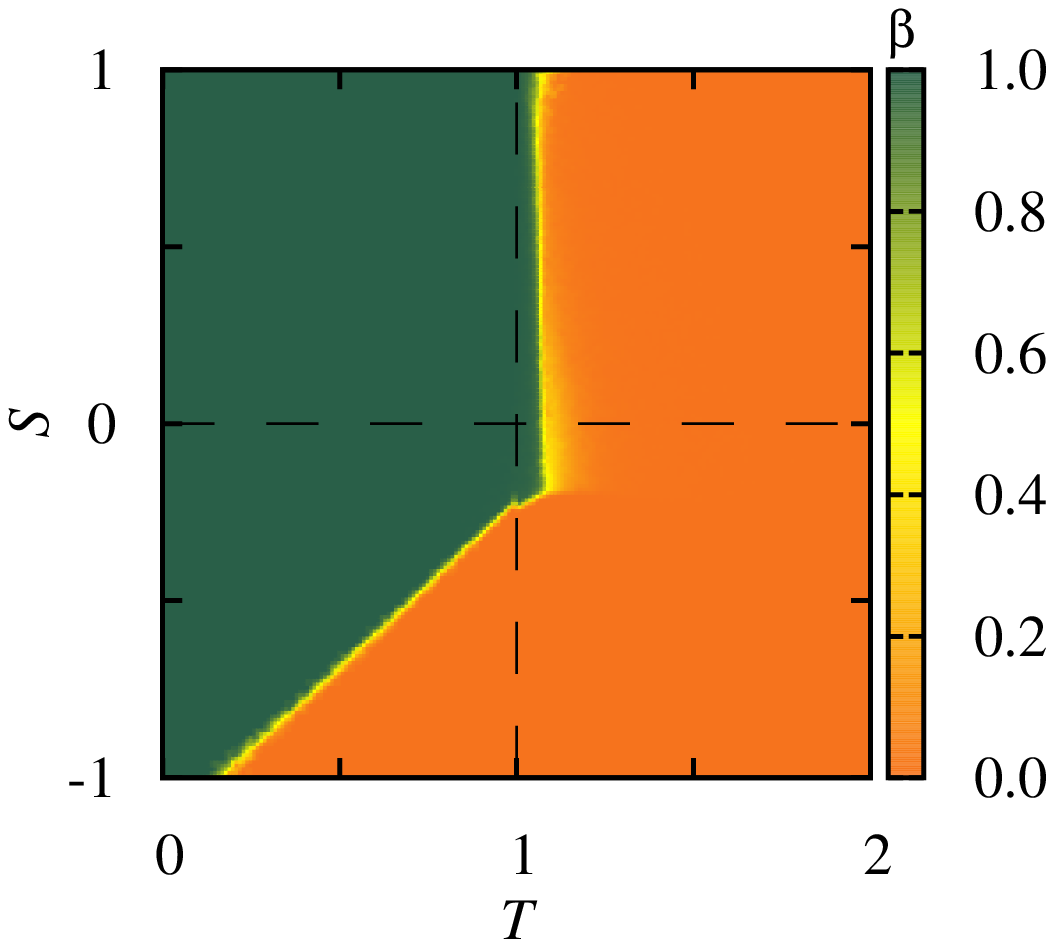,width=4.3cm}}
\caption{\label{RRG} Color map depicting the final values of $\alpha$ (left) and $\beta$ (right) on the $T-S$ parameter plane, as obtained on a regular random graph. The results are strikingly similar as obtained on a square lattice (see \cite{szolnoki_epl11}).}
\end{figure}

\begin{figure}[b]
\centerline{\epsfig{file=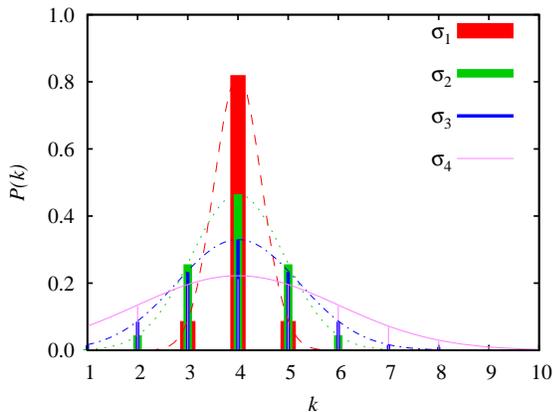,width=8cm}}
\caption{\label{dist} Degree distributions of random graphs with a gradually increasing variance of degree ($\sigma_1 < \sigma_2 < \sigma 3 < \sigma_4$). For easier reference the envelopes of discrete degree distributions are depicted as well.}
\end{figure}

We start by presenting results obtained on a regular random graph \cite{szabo_jpa04} with $k=4$, as it is a natural extension of a simple square lattice population which we have considered before in \cite{szolnoki_epl11}. Importantly, while the degree distribution remains uniform, other topological features, like the presence of shortcuts or the emergence of a nonzero clustering coefficient, change significantly. Previous works on games using pure strategies highlighted that these details may play a significant role by the evolution of cooperation in social dilemmas \cite{szabo_pre05, vukov_pre06, vilone_jsm11}. Figure~\ref{RRG} depicts the color map encoding the fixation values of $\alpha$ (left) and $\beta$ (right) on the $T-S$ parameter plane. From the presented results it follows that if the governing social dilemma is of the snowdrift type, players will always (never) cooperate with their neighbors provided their payoff is lower (higher). In the prisoner's dilemma quadrant, we can observe either complete dominance of defection regardless of the status of the opponents, or the same situation as in the snowdrift quadrant provided $S$ is not too negative. For the stag-hunt and the harmony game the outcome is practically identical as obtained by means of the traditional version of the two games. In general, however, both color profiles differ only insignificantly from the ones we have reported in \cite{szolnoki_epl11} (see Figs.~2 and 3 there) for the square lattice. This leads to the conclusion that the structure of interactions does not play a prominent role as long as the degree of all players is uniform.

This leads to suspect that the heterogeneity of interactions might play a pivotal role. We therefore depart from the regular random graph to random graphs with different degree distributions, as depicted in Fig.~\ref{dist}. We consider four different types of random
graphs with Gaussian distributed degree, yet with increasing variance. According to the legend of Fig.~\ref{dist}, the random graph with $\sigma_1$ is thus the least heterogeneous (only degrees $k=3$, $4$ and $5$ are possible), while the random graph with $\sigma_4$ is the most heterogeneous. Increasing gradually the variance from $\sigma_1$ to $\sigma_4$ thus enables us to monitor directly the consequences of heterogeneity stemming from the interaction network.

\begin{figure}
\centerline{\epsfig{file=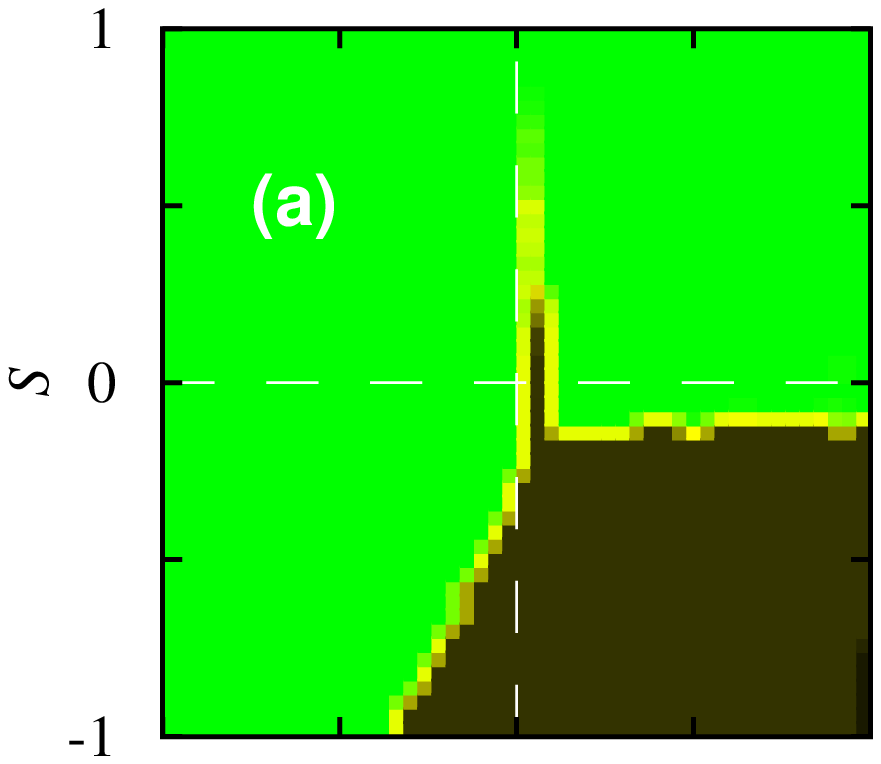,width=2.1cm}\epsfig{file=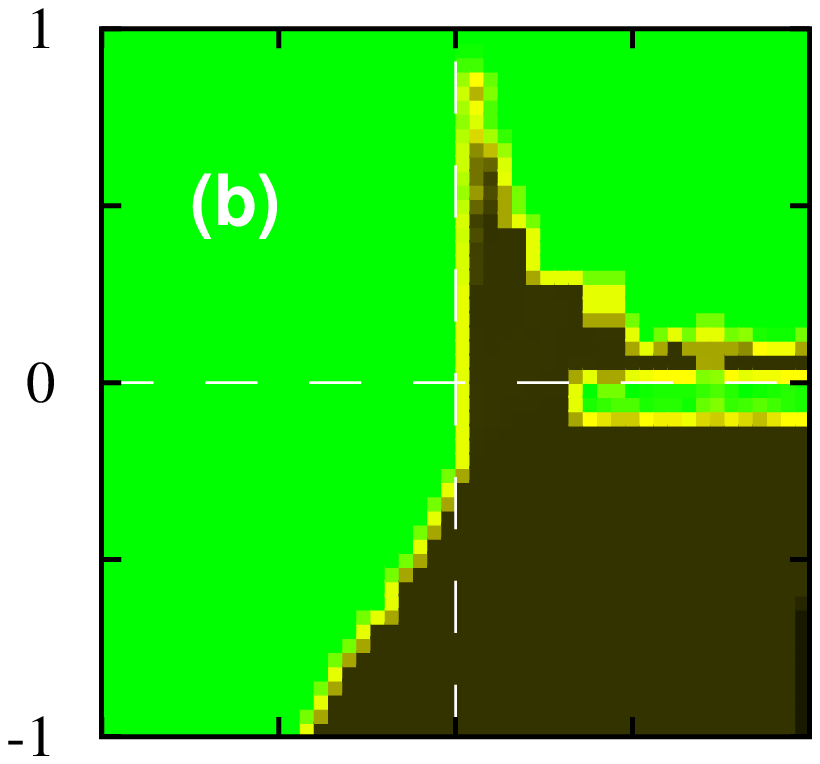,width=2.1cm}\epsfig{file=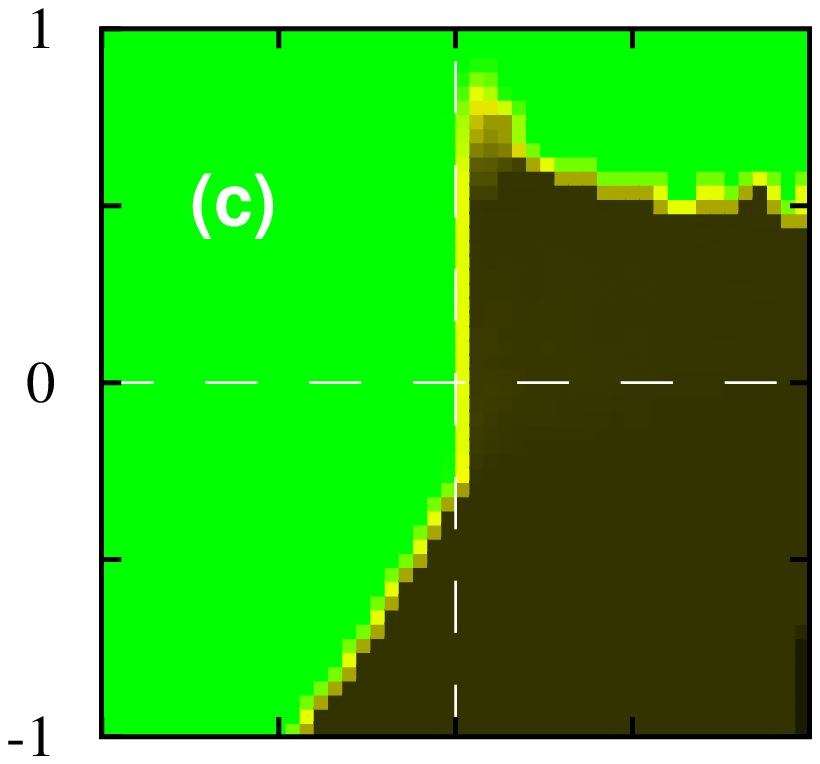,width=2.1cm}\epsfig{file=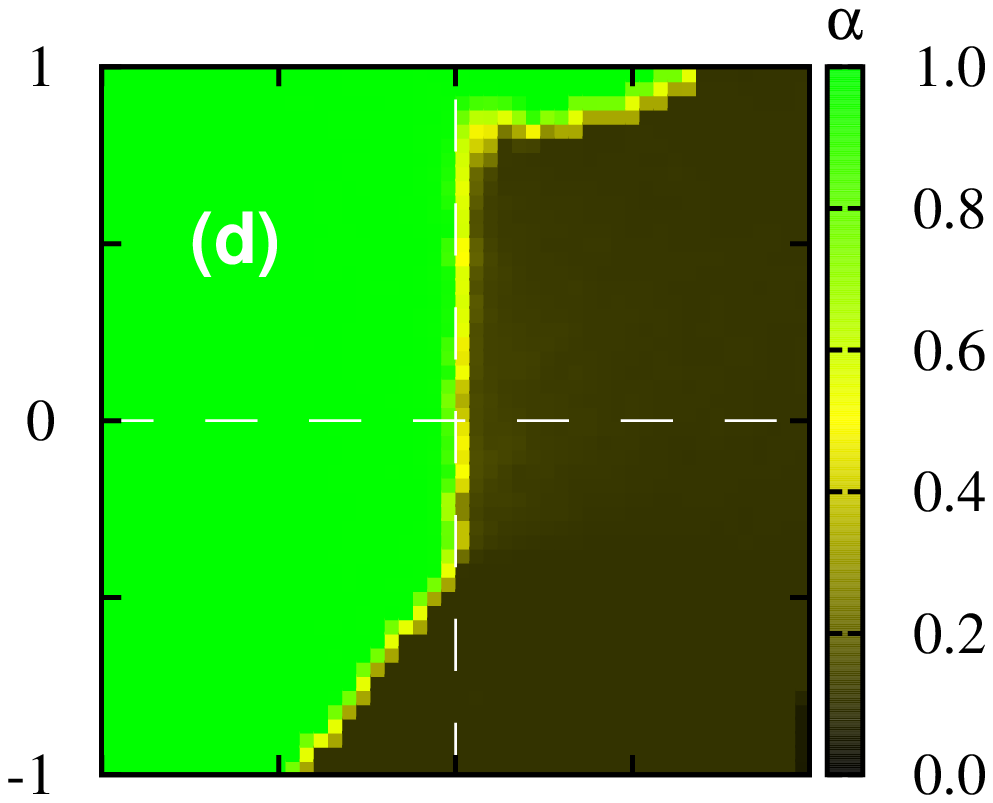,width=2.1cm}}
\centerline{\epsfig{file=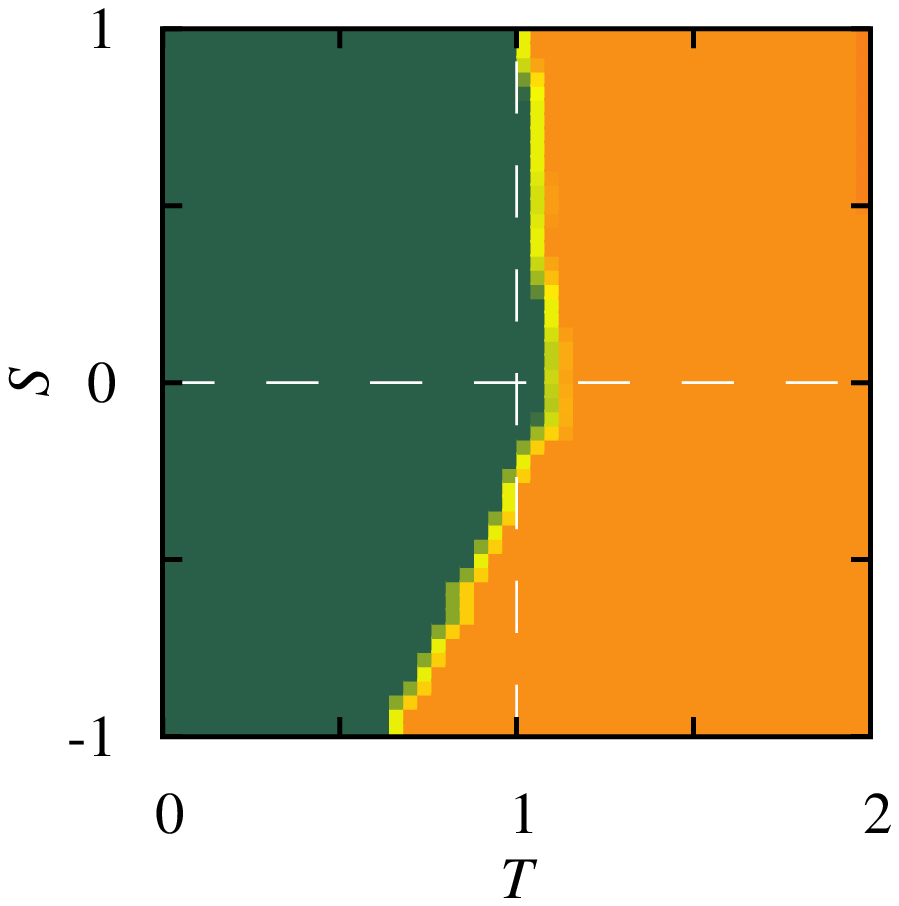,width=2.1cm}\epsfig{file=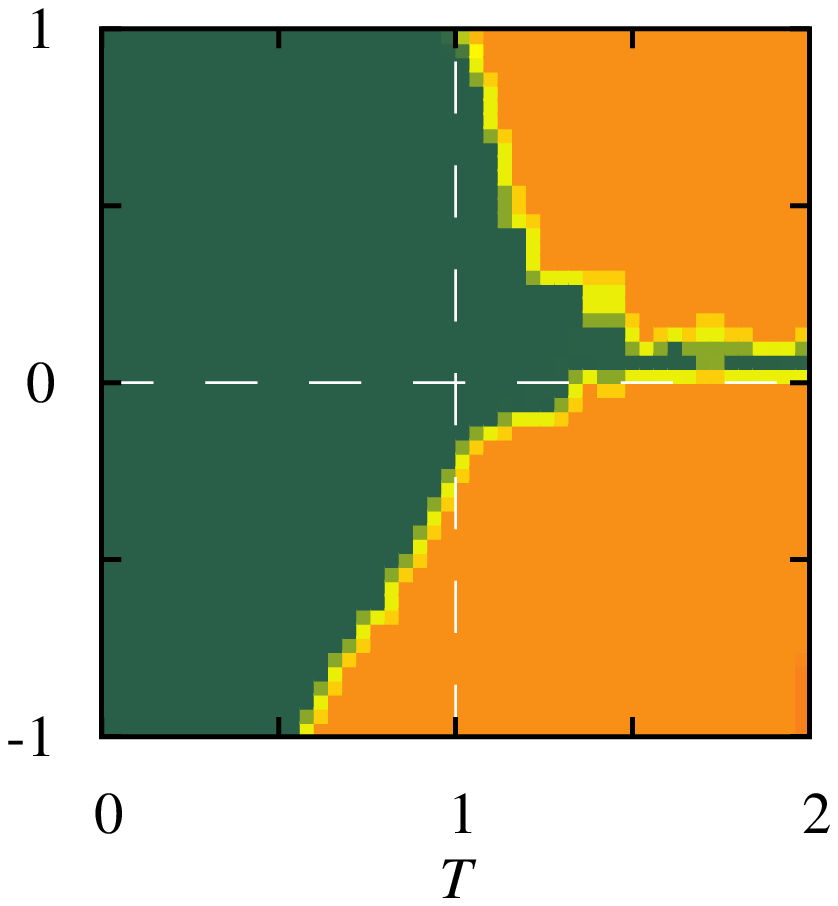,width=2.1cm}\epsfig{file=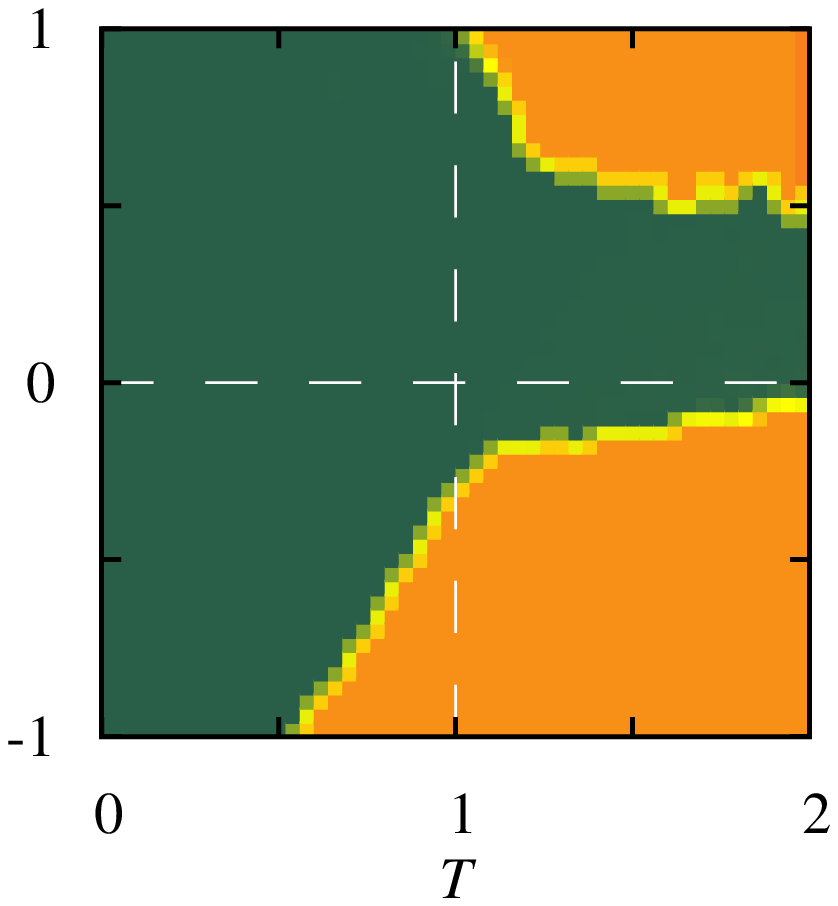,width=2.1cm}\epsfig{file=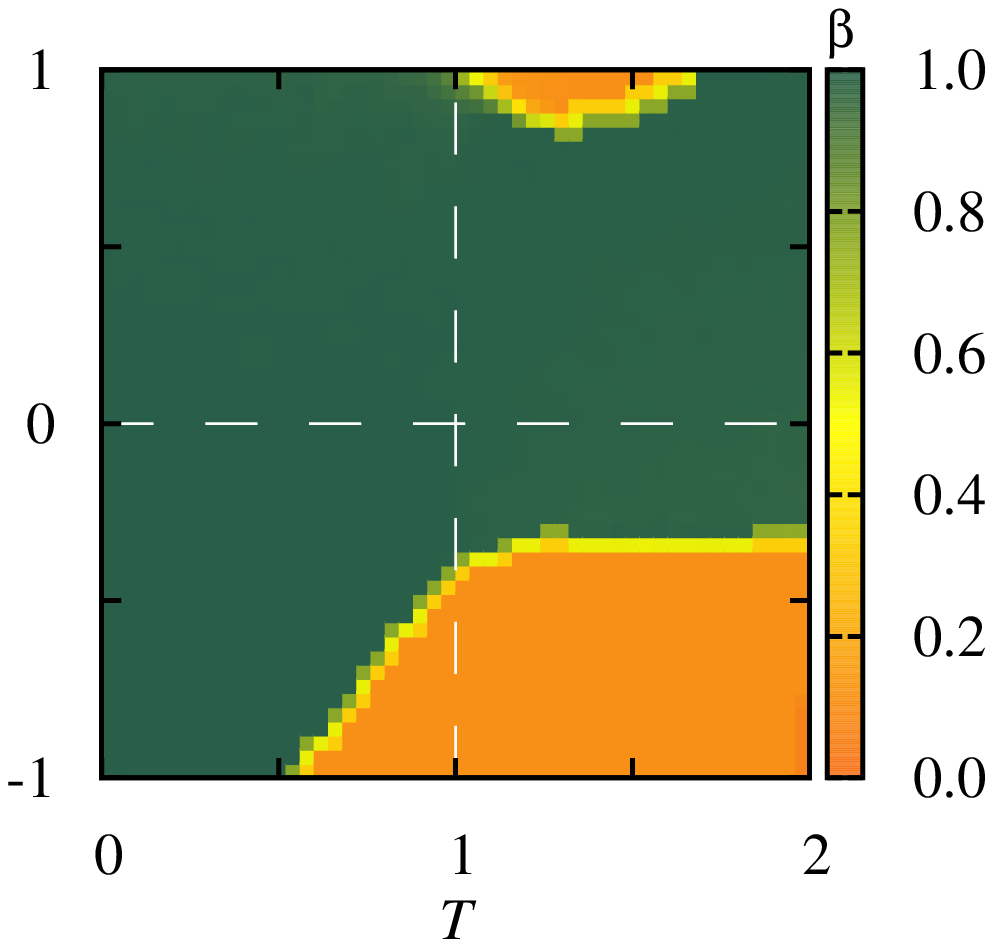,width=2.1cm}}
\caption{\label{randoms} Color maps depicting the final values of $\alpha$ (top) and $\beta$ (bottom) on the $T-S$ parameter plane, as obtained on random graphs with degree distribution as depicted in Fig.~\ref{dist}. From left to right the variance increases from $\sigma_1$ to $\sigma_2$ to $\sigma_3$ to $\sigma_4$, and hence increases also the network heterogeneity. It can be observed that the higher the heterogeneity, the more the high-$\alpha$ low-$\beta$ emotional profiles give way to profiles that are characterized by low-$\alpha$ and high-$\beta$ values. The transition is particularly pronounced in the snowdrift and the prisoner's dilemma quadrant, while for the stag-hunt and the harmony game the outcome remains little affected.}
\end{figure}

Color maps encoding the fixation values of $\alpha$ and $\beta$ for the four different random graphs are depicted in Fig.~\ref{randoms}. By following the plots from left to right,
it can be observed that as the heterogeneity of the interaction network increases, the fixation of profiles of both $\alpha$ and $\beta$ change. By focusing on the snowdrift and the prisoner's dilemma quadrant, there is a gradual shift from high-$\alpha$ low-$\beta$ emotional profiles to low-$\alpha$ and high-$\beta$ values as heterogeneity increases. Accordingly, taking into account also results presented in Fig.~\ref{RRG}, we conclude that homogeneous interaction networks promote emotions like sympathy and envy ($\alpha \to 1$ and $\beta \to 0$), while heterogeneous interaction network prefer indifference and servility ($\alpha \to 0$ and $\beta \to 1$). It is worth highlighting that these emotional profiles emerge completely spontaneously based on payoff-driven imitation. The change is thus brought about exclusively by the heterogeneity of the interaction network.

\begin{figure}
\centerline{\epsfig{file=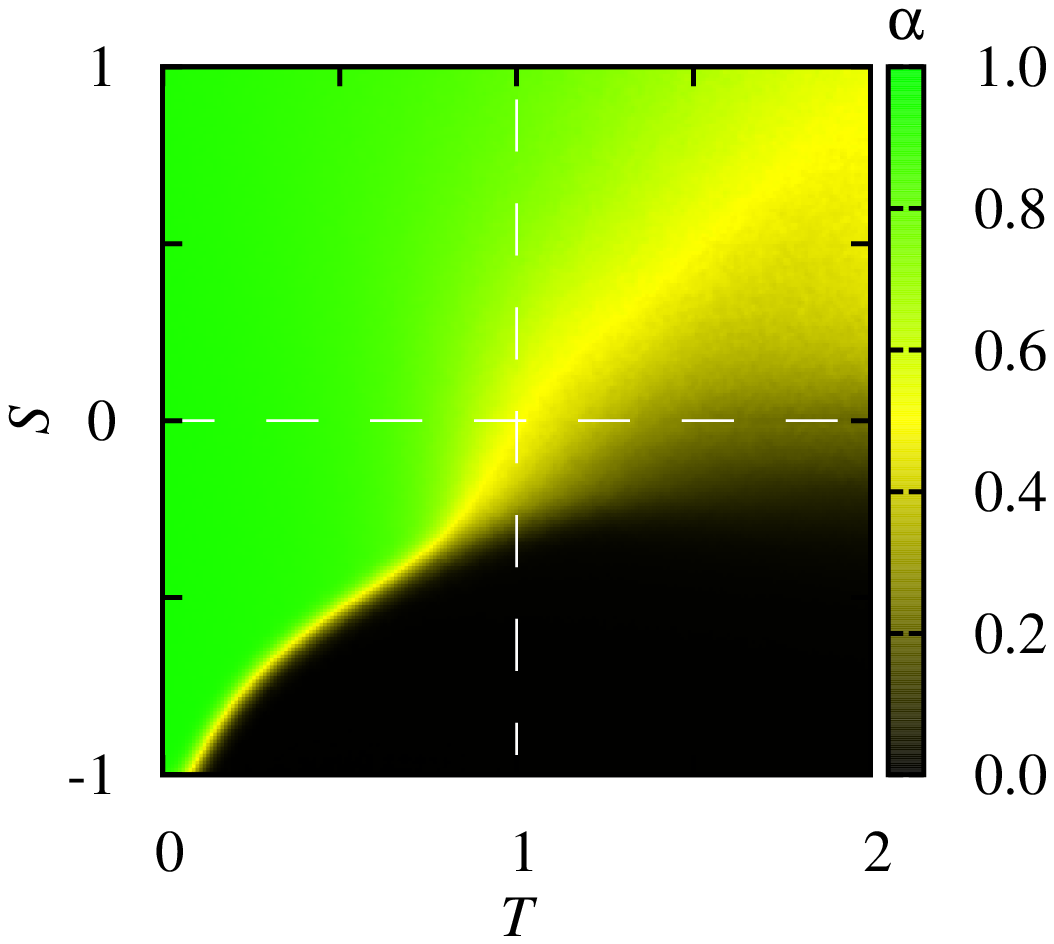,width=4.3cm}\epsfig{file=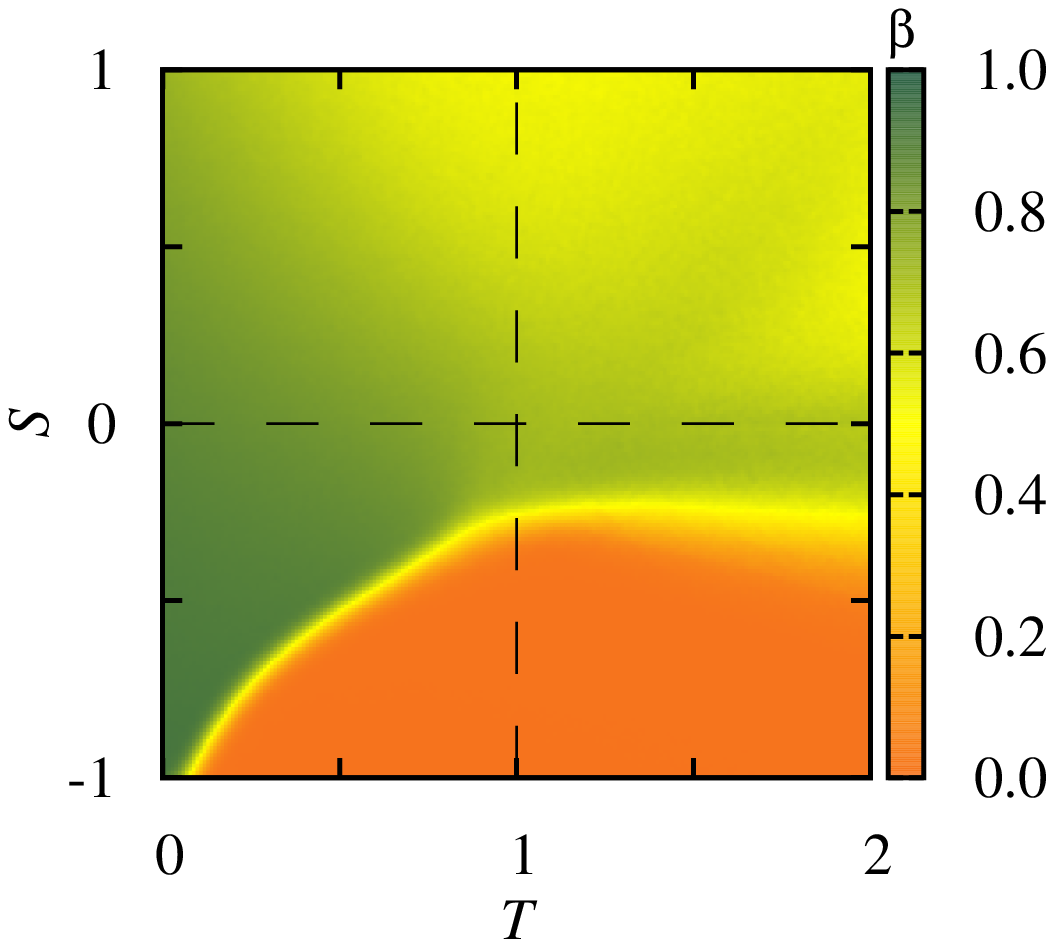,width=4.3cm}}
\caption{\label{SF} Color map depicting the final values of $\alpha$ (left) and $\beta$ (right) on the $T-S$ parameter plane, as obtained on a scale-free network. The dominance of low and moderate $\alpha$ values and high $\beta$ values is even more pronounced than for the random graph with degree distribution $\sigma_4$ (compare with the two rightmost panels in Fig.~\ref{randoms}). This further strengthens the conclusion that, unlike homogeneous networks, strong heterogeneity strongly favors the fixation of emotional profiles that are characterized by low-$\alpha$ and high-$\beta$ values.}
\end{figure}

It is possible to take a step further in terms of the heterogeneity of the interaction network by considering scale-free networks. We therefore make use of the standard model proposed by Barab{\'a}si and Albert \cite{barabasi_s99}. Results presented in Fig.~\ref{SF} further support our arguments, as the region of low and moderate $\alpha$ values extends further into the snowdrift quadrant, while at the same time low $\beta$ values vanish more and more from both the snowdrift and the prisoner's dilemma quadrant. As before, the harmony games and the stag-hunt quadrant remain relatively unaffected, which corroborates the fact that the proposed shift from the imitation of strategies to the imitation of emotional profiles affect predominantly the social dilemma games. It is also worth reminding that on scale-free networks the fixation may not be unique because different hubs can sustain their own micro-environment independently from the other hubs. We therefore depict an average over several independent realizations to arrive at representative results.

\begin{figure}
\centerline{\epsfig{file=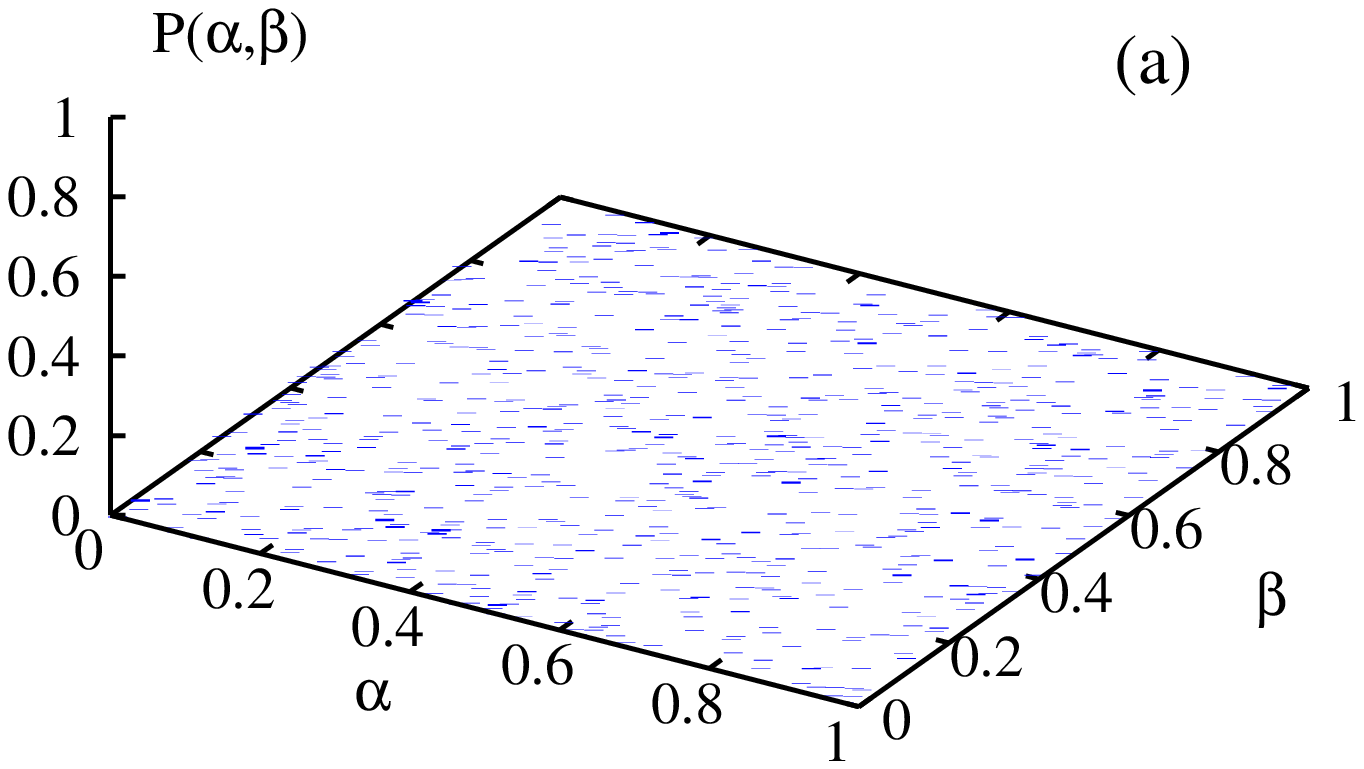,width=4.3cm}\epsfig{file=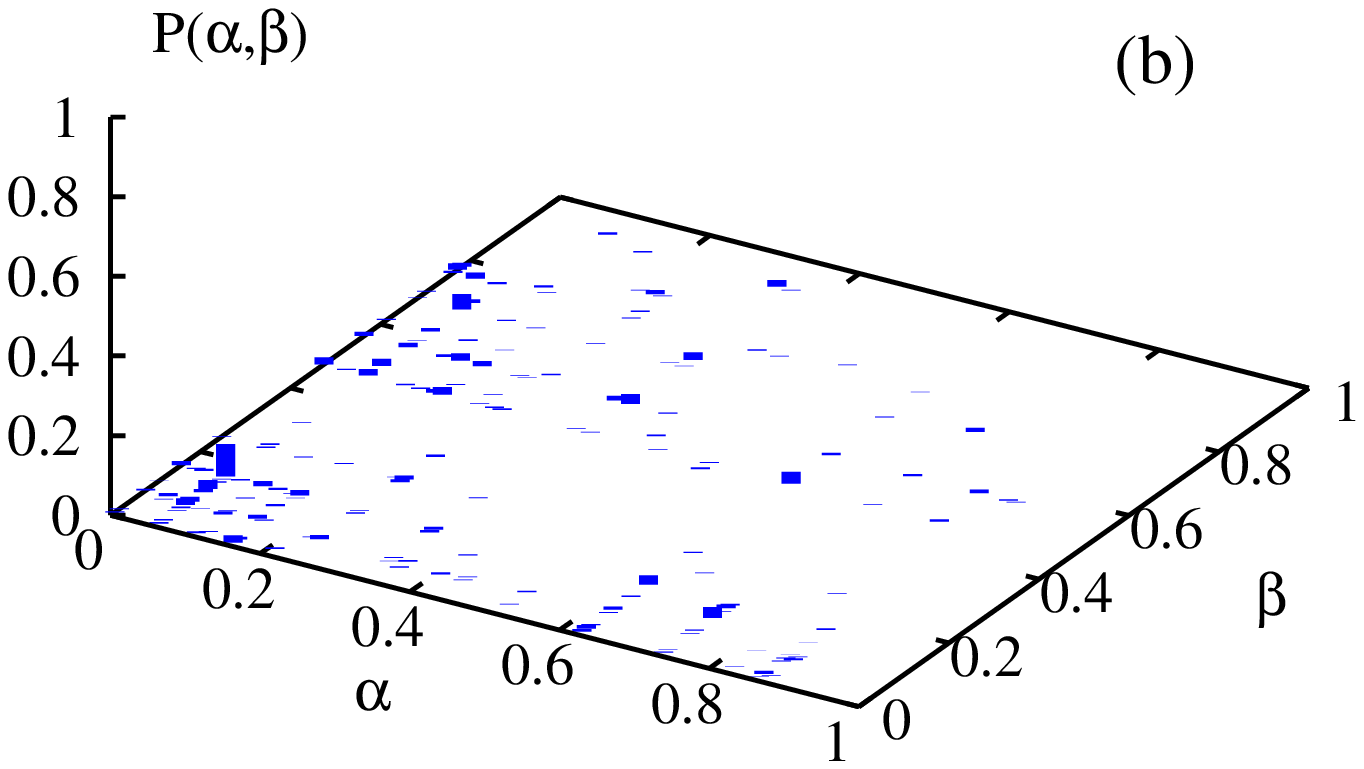,width=4.3cm}}
\centerline{\epsfig{file=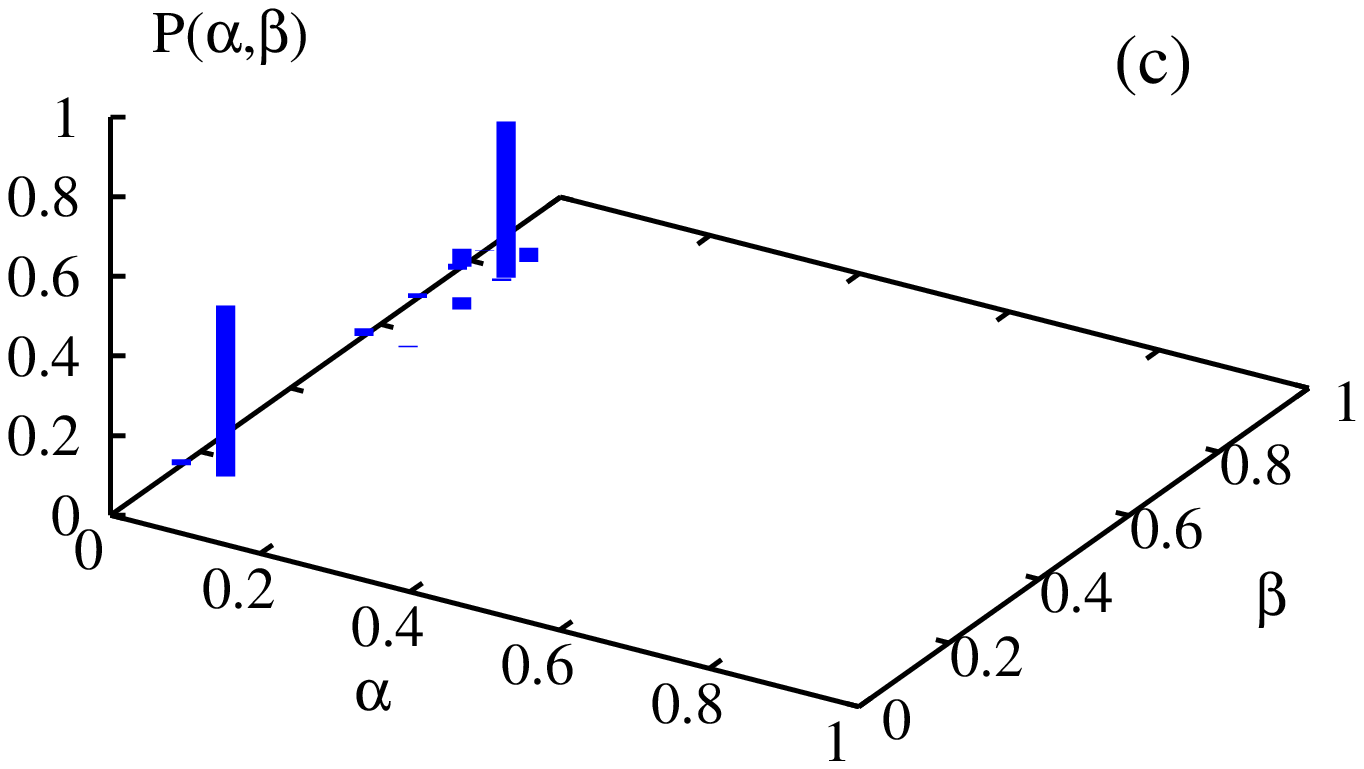,width=4.3cm}\epsfig{file=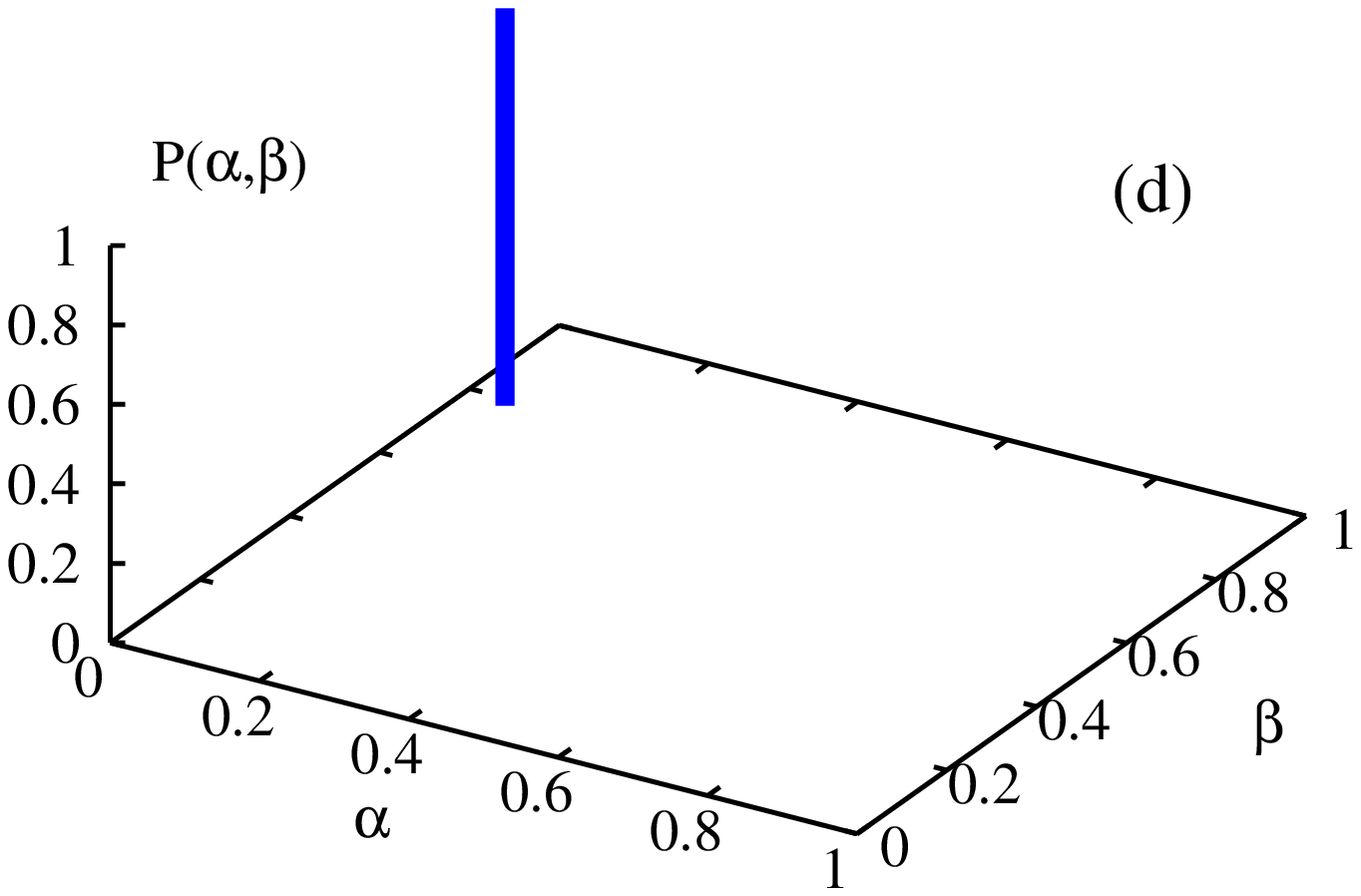,width=4.3cm}}
\caption{\label{fixation} Time evolution of fixation of $\alpha$ and $\beta$, as obtained for $T=1.5$ and $S=-0.1$ on the scale-free network. From top left to bottom right we have the temporary distribution of $(\alpha_x, \beta_x)$ pairs at $1$, $100$, $1000$ and $100000$ full Monte Carlo steps using $N=5000$ players. It can be observed that high values of $\alpha$ are the first to vanish. Gradually then also the remaining low $\beta$ values give way to the complete dominance of low-$\alpha$ and high-$\beta$ emotional profiles.}
\end{figure}

In order to obtain an understanding of the preference for low-$\alpha$ and high-$\beta$ values, as it is exerted by heterogeneous interaction networks, it is of interest to examine the time evolution of $\alpha$ and $\beta$ values, as depicted in Fig.~\ref{fixation}. The figure shows the probability of any given $(\alpha, \beta)$ pair in the population at different times increasing from top left to bottom right. It can be observed that high-$\alpha$ -- high-$\beta$ combinations die out first. These players cooperate with both their more and less successful opponents, and they do so with a high probability. In agreement with well-known results concerning the evolution of cooperation in spatial social dilemmas \cite{szabo_pr07}, the bulk of cooperators is always the first to die out. Only after their arrangement in suitable compact domains the cooperators can take advantage of network reciprocity and prevail against defectors. In our case, however, this does not happen, i.e., the ``always cooperate'' players never recover. Instead, the evolution proceeds by eliminating also all pairs which contain moderate and high $\alpha$ values, until finally the only surviving low-$\alpha$ profiles are left to compete. However, preserving at least some form of cooperation may yield an evolutionary advantage, and thus ultimately the low-$\alpha$ -- high-$\beta$ emotional profile emerges as the only remaining. Notably, the described scenario is characteristic only for heterogeneous networks. For homogeneous networks the differences between players are more subtle, and indeed it is not at all obvious that cooperating with the more successful neighbors would confer an evolutionary advantage. Accordingly, high-$\beta$ profiles are not viable and die out. Cooperation can thrive only on the expense of high $\alpha$ values, as reported already in \cite{szolnoki_epl11}.

\begin{figure}
\centerline{\epsfig{file=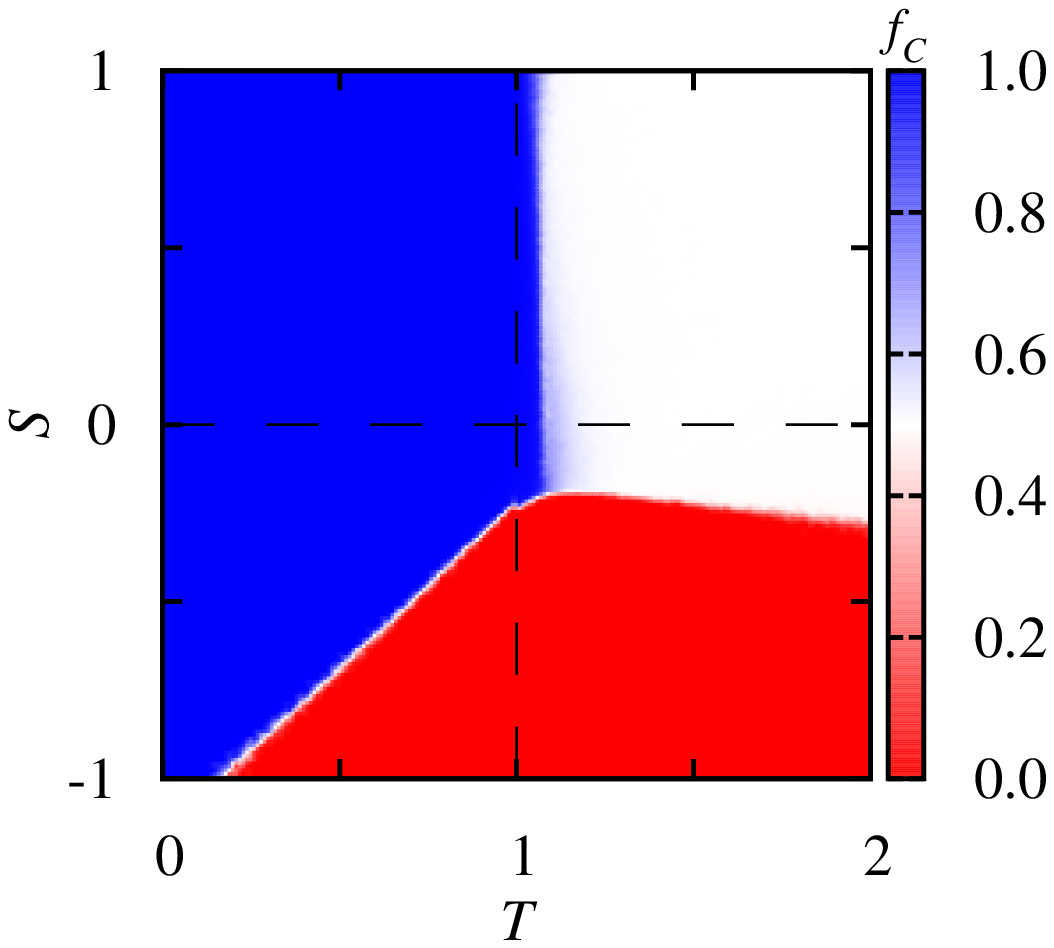,width=4.3cm}\epsfig{file=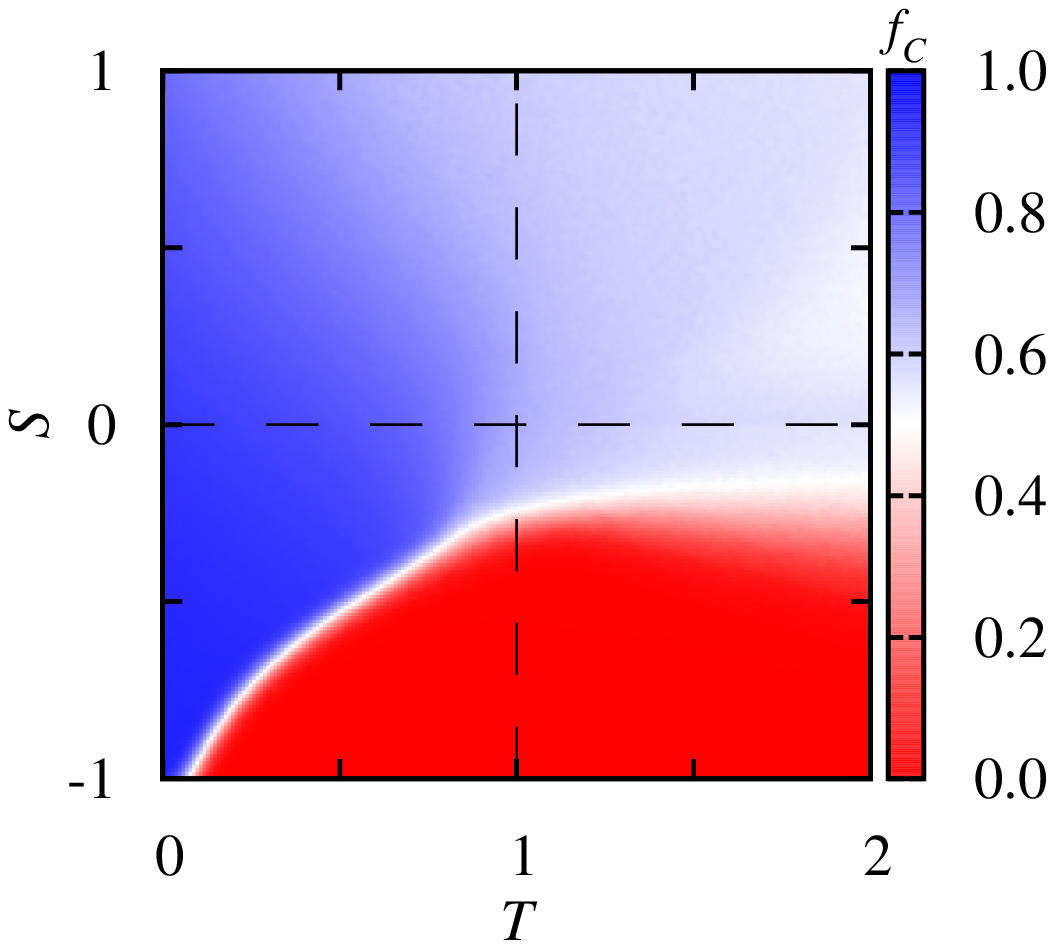,width=4.3cm}}
\caption{\label{cooperate} Color map depicting the final probability of cooperative behavior $f_C$ on the $T-S$ parameter plane, as obtained on the regular random graph (left) and the scale-free network (right). Since the probability to cooperate should be seen equal to the stationary fraction of cooperators in the traditional version of the game, a comparison of presented results (compare with Fig.~1 in \cite{szolnoki_epl11}) reveals that replacing the imitation of strategies with the imitation of emotional profiles strongly promotes the evolution of cooperation. Even more so if the interaction network is strongly heterogeneous.}
\end{figure}

Importantly though, given an appropriately heterogeneous interaction network, the low-$\alpha$ -- high-$\beta$ emotional profile can be very much beneficial for the global cooperation level. To support this statement, we present in Fig.~\ref{cooperate} the average frequency of cooperation as obtained on the regular random graph (left) and the scale-free network (right). Note that the former in general represent homogeneous graphs. The comparison reveals that a much higher cooperation level can be sustained, especially in the snowdrift quadrant, if the dominating emotions are neither sympathy nor envy. To confirm this further, we have manually imposed a high-$\alpha$ -- low-$\beta$ emotional profile on the scale-free network. While this profile is optimal for homogeneous networks (compare also Fig.~\ref{cooperate} (left) with Fig.~4 in \cite{szolnoki_epl11}) the outcome on heterogeneous networks is disappointing, yielding no more than a modest cooperation level of $f_C \approx 0.30-0.35$ in the most challenging snowdrift and prisoner's dilemma regions. This imposes another interesting conclusion, namely, if the emotional profiles of players can evolve freely as dictated by payoff-driven imitation microscopic dynamics, then the topology ``selects'' the optimal profile in order to produce the highest attainable cooperation level.

\begin{figure}
\centerline{\epsfig{file=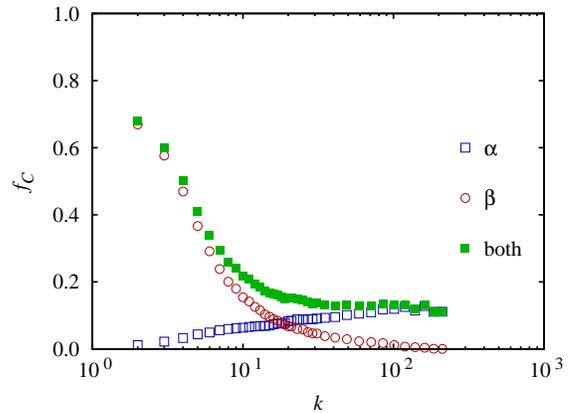,width=8cm}}
\caption{\label{degree} The average willingness to cooperate in dependence on degree, as obtained for $T=1.5$ and $S=-0.1$ on the scale-free network. Fully in agreement with the dominant low-$\alpha$ and high-$\beta$ emotional profile, it can be observed that hubs very rarely cooperate, while the masses do so almost always. Depicted result is an average over 500 independent runs at $N=5000$ after $10^7$ Monte Carlo steps. The average cooperation level is $\approx 0.6$.}
\end{figure}

Lastly, it is instructive to explore how the low-$\alpha$ -- high-$\beta$ emotional profile actually works on scale-free networks. A visualization is possible by measuring separately the average willingness to cooperate for players who have different degree. Since the payoff of every player is obtained from the pairwise interaction constituted by each individual link, a higher degree $k$ therefore in general leads to a higher payoff and also a higher ``social prestige''. As Fig.~\ref{degree} illustrates, players with low degree will dominantly cooperate with their opponents that have a higher degree and thus most likely a higher payoff. In other words, they can use the ``$\beta$-part'' of their emotional profile. This act of cooperation, however, is unilateral because the hubs rarely compensate it. Due to low values of $\alpha$ cooperation with the less successful players is strongly suppressed. What is more, while players with a higher degree also cooperate with the more successful opponents (they have the same emotional profile and hence the same high $\beta$), this action is very rare given that there are simply not many who would be superior. It is sad but still true that the hubs with the highest degree very rarely cooperate in the stationary state. Despite this rather unfriendly behavior of the ``leaders'', the average cooperation level is still acceptable and in fact remarkably high even under adverse conditions (e.g., $T=1.5$ and $S=-0.1$), but this is exclusively because the inferior players do their best and virtually always cooperate towards their superiors.

\section{Discussion}
We have shown that high levels of cooperation can evolve amongst self-interested individuals if instead of strategies they adopt simple emotional profiles from their neighbors. Since the imitation was governed solely by the payoffs of players, we have made no additional assumptions concerning the microscopic dynamics. The later has been governed by the traditional ``follow the more successful'' rule, which we have implemented with some leeway due to the Fermi function. Starting from an initial configuration with all possible emotional profiles, we have determined the one that remains after sufficiently long relaxation (only in the harmony game quadrant, if staged on heterogeneous networks, the fixation may not be unique). We have found that the fixation depends not only on the parametrization of the game, but even more so on the topology of the interaction network. More precisely, the topology-induced heterogeneity of players has been identified as the most important property. If players were staged on a network where their degree was equal, then independently of other topological properties of the network the fixation occurred on emotional profiles characterized by high $\alpha$ and low $\beta$ values in the interesting payoff region. In agreement with the definition of $\alpha$ and $\beta$, these are players characterized by high sympathy but also high envy. This profile is also in agreement with the one reported earlier in \cite{szolnoki_epl11} for the square lattice. On heterogeneous networks, however, the fixation is most likely to be on low or moderate values of $\alpha$ and high values of $\beta$. Accordingly, we have the prevalence of low sympathy (charity, goodwill) for those that are doing worse, but also little envy (high servility, proneness to brownnose or ``suck up'') towards those that are doing better. Noteworthy, although we have not presented actual results, the application of payoffs normalized with the degree of players returns the same results as observed on homogeneous networks. This observation is in agreement with our preliminary expectation because it is well established that the scale-free topology introduces a strong heterogeneity amongst players, but also that this effect is effectively diminished by applying normalized payoffs or degree-sensitive cost \cite{santos_jeb06, tomassini_ijmpc07, wu_zx_pa07, masuda_prsb07, szolnoki_pa08}. Accordingly, in the latter case players become ``equal'', which results in the selection of emotional profile we have recorded for regular graphs. This observation strengthens further our argument that indeed solely the heterogeneity of players is crucial for the selection of the dominant emotional profile.

We thus may argue that on heterogeneous networks each ``dictators' dream'' profile can evolve via a simple evolutionary rule. The majority may not be happy about it because the combination of moderate $\alpha$ and high $\beta$ values is not necessarily the most coveted personality profile. Yet as our study shows, it does have its social advantages. Namely, in the absence of envy or in the presence of servility the cooperation level in the whole population can be maintained relatively very high, even if the conditions for the evolution of cooperation are extremely adverse (high $T$, low $S$). In this sense, we conclude that charity and envy are easily outperformed by competitiveness and proneness to please the dominant players, and that indeed this profile emerges completely spontaneous. Put differently, it can be argued that it is in fact chosen by the heterogeneity amongst players that is introduced by an appropriate interaction network.

We would also like to emphasize that the discussed ``emotional profiles'' do not necessarily cover the broader psychological interpretation of the term \cite{davidson_12}. We have used this terminology to express the liberty of each individual to act differently towards different partners in dependence on the differences in social rank (or success), which traditional strategies in the context of evolutionary games do not allow. As such, and in the absence of considering further details determining our personality, our very simple model naturally cannot be held accountable for describing actual human behavior. Instead, it reveals the topology of interactions as a crucial property that determines the collective behavior of a social network. According to our observations, it is indeed the heterogeneity of the interaction network that is key in determining our willingness to help others.

Finally, we emphasize that in the present model cooperation is maintained without reciprocity. The mechanism at work here is very different from those discussed thoroughly in previous studies. Unlike direct and indirect reciprocity, network reciprocity, or even reputation, punishment and reward, which are all deeply routed in the fact that neighboring cooperators will help each other out while at the same time neighboring defectors will craft their own demise, here the nature of links determines the winner. It may well happen that cooperation and defection occur along the same link, yet still the status of the population as a whole is very robust. What players really share is the way how to behave towards each other under different circumstances, which is determined within the framework of an emotional profile.\\ \\

\begin{acknowledgments}
This research was supported by the Hungarian National Research Fund (grant K-101490), China's Education Ministry via the Humanities and Social Sciences research project (11YJC630208), and the Slovenian Research Agency (grant J1-4055).
\end{acknowledgments}


\begin{thebibliography}{56}
\expandafter\ifx\csname natexlab\endcsname\relax\def\natexlab#1{#1}\fi
\expandafter\ifx\csname bibnamefont\endcsname\relax
  \def\bibnamefont#1{#1}\fi
\expandafter\ifx\csname bibfnamefont\endcsname\relax
  \def\bibfnamefont#1{#1}\fi
\expandafter\ifx\csname citenamefont\endcsname\relax
  \def\citenamefont#1{#1}\fi
\expandafter\ifx\csname url\endcsname\relax
  \def\url#1{\texttt{#1}}\fi
\expandafter\ifx\csname urlprefix\endcsname\relax\def\urlprefix{URL }\fi
\providecommand{\bibinfo}[2]{#2}
\providecommand{\eprint}[2][]{\url{#2}}

\bibitem[{\citenamefont{Hofbauer and Sigmund}(1998)}]{hofbauer_98}
\bibinfo{author}{\bibfnamefont{J.}~\bibnamefont{Hofbauer}} \bibnamefont{and}
  \bibinfo{author}{\bibfnamefont{K.}~\bibnamefont{Sigmund}},
  \emph{\bibinfo{title}{Evolutionary Games and Population Dynamics}}
  (\bibinfo{publisher}{Cambridge University Press},
  \bibinfo{address}{Cambridge, U.K.}, \bibinfo{year}{1998}).

\bibitem[{\citenamefont{Liggett}(1985)}]{liggett_85}
\bibinfo{author}{\bibfnamefont{T.~M.} \bibnamefont{Liggett}},
  \emph{\bibinfo{title}{Interacting Particle Systems}}
  (\bibinfo{publisher}{Springer}, \bibinfo{address}{New York},
  \bibinfo{year}{1985}).

\bibitem[{\citenamefont{Castellano et~al.}(2009)\citenamefont{Castellano,
  Fortunato, and Loreto}}]{castellano_rmp09}
\bibinfo{author}{\bibfnamefont{C.}~\bibnamefont{Castellano}},
  \bibinfo{author}{\bibfnamefont{S.}~\bibnamefont{Fortunato}},
  \bibnamefont{and} \bibinfo{author}{\bibfnamefont{V.}~\bibnamefont{Loreto}},
  \bibinfo{journal}{Rev. Mod. Phys.} \textbf{\bibinfo{volume}{81}},
  \bibinfo{pages}{591} (\bibinfo{year}{2009}).

\bibitem[{\citenamefont{Axelrod}(1984)}]{axelrod_84}
\bibinfo{author}{\bibfnamefont{R.}~\bibnamefont{Axelrod}},
  \emph{\bibinfo{title}{The Evolution of Cooperation}}
  (\bibinfo{publisher}{Basic Books}, \bibinfo{address}{New York},
  \bibinfo{year}{1984}).

\bibitem[{\citenamefont{Macy and Flache}(2002)}]{macy_pnas02}
\bibinfo{author}{\bibfnamefont{M.~W.} \bibnamefont{Macy}} \bibnamefont{and}
  \bibinfo{author}{\bibfnamefont{A.}~\bibnamefont{Flache}},
  \bibinfo{journal}{Proc. Natl. Acad. Sci. USA} \textbf{\bibinfo{volume}{99}},
  \bibinfo{pages}{7229} (\bibinfo{year}{2002}).

\bibitem[{\citenamefont{Hardin}(1968)}]{hardin_g_s68}
\bibinfo{author}{\bibfnamefont{G.}~\bibnamefont{Hardin}},
  \bibinfo{journal}{Science} \textbf{\bibinfo{volume}{162}},
  \bibinfo{pages}{1243} (\bibinfo{year}{1968}).

\bibitem[{\citenamefont{Nowak}(2006)}]{nowak_s06}
\bibinfo{author}{\bibfnamefont{M.~A.} \bibnamefont{Nowak}},
  \bibinfo{journal}{Science} \textbf{\bibinfo{volume}{314}},
  \bibinfo{pages}{1560} (\bibinfo{year}{2006}).

\bibitem[{\citenamefont{Szab{\'o} and F{\'a}th}(2007)}]{szabo_pr07}
\bibinfo{author}{\bibfnamefont{G.}~\bibnamefont{Szab{\'o}}} \bibnamefont{and}
  \bibinfo{author}{\bibfnamefont{G.}~\bibnamefont{F{\'a}th}},
  \bibinfo{journal}{Phys. Rep.} \textbf{\bibinfo{volume}{446}},
  \bibinfo{pages}{97} (\bibinfo{year}{2007}).

\bibitem[{\citenamefont{Schuster et~al.}(2008)\citenamefont{Schuster, Kreft,
  Schroeter, and Pfeiffer}}]{schuster_jbp08}
\bibinfo{author}{\bibfnamefont{S.}~\bibnamefont{Schuster}},
  \bibinfo{author}{\bibfnamefont{J.-U.} \bibnamefont{Kreft}},
  \bibinfo{author}{\bibfnamefont{A.}~\bibnamefont{Schroeter}},
  \bibnamefont{and} \bibinfo{author}{\bibfnamefont{T.}~\bibnamefont{Pfeiffer}},
  \bibinfo{journal}{J. Biol. Phys.} \textbf{\bibinfo{volume}{34}},
  \bibinfo{pages}{1} (\bibinfo{year}{2008}).

\bibitem[{\citenamefont{Roca et~al.}(2009)\citenamefont{Roca, Cuesta, and
  S{\'a}nchez}}]{roca_plr09}
\bibinfo{author}{\bibfnamefont{C.~P.} \bibnamefont{Roca}},
  \bibinfo{author}{\bibfnamefont{J.~A.} \bibnamefont{Cuesta}},
  \bibnamefont{and}
  \bibinfo{author}{\bibfnamefont{A.}~\bibnamefont{S{\'a}nchez}},
  \bibinfo{journal}{Phys. Life Rev.} \textbf{\bibinfo{volume}{6}},
  \bibinfo{pages}{208} (\bibinfo{year}{2009}).

\bibitem[{\citenamefont{Perc and Szolnoki}(2010)}]{perc_bs10}
\bibinfo{author}{\bibfnamefont{M.}~\bibnamefont{Perc}} \bibnamefont{and}
  \bibinfo{author}{\bibfnamefont{A.}~\bibnamefont{Szolnoki}},
  \bibinfo{journal}{BioSystems} \textbf{\bibinfo{volume}{99}},
  \bibinfo{pages}{109} (\bibinfo{year}{2010}).

\bibitem[{\citenamefont{Nowak and May}(1992)}]{nowak_n92b}
\bibinfo{author}{\bibfnamefont{M.~A.} \bibnamefont{Nowak}} \bibnamefont{and}
  \bibinfo{author}{\bibfnamefont{R.~M.} \bibnamefont{May}},
  \bibinfo{journal}{Nature} \textbf{\bibinfo{volume}{359}},
  \bibinfo{pages}{826} (\bibinfo{year}{1992}).

\bibitem[{\citenamefont{Santos and Pacheco}(2005)}]{santos_prl05}
\bibinfo{author}{\bibfnamefont{F.~C.} \bibnamefont{Santos}} \bibnamefont{and}
  \bibinfo{author}{\bibfnamefont{J.~M.} \bibnamefont{Pacheco}},
  \bibinfo{journal}{Phys. Rev. Lett.} \textbf{\bibinfo{volume}{95}},
  \bibinfo{pages}{098104} (\bibinfo{year}{2005}).

\bibitem[{\citenamefont{Santos et~al.}(2006)\citenamefont{Santos, Pacheco, and
  Lenaerts}}]{santos_pnas06}
\bibinfo{author}{\bibfnamefont{F.~C.} \bibnamefont{Santos}},
  \bibinfo{author}{\bibfnamefont{J.~M.} \bibnamefont{Pacheco}},
  \bibnamefont{and} \bibinfo{author}{\bibfnamefont{T.}~\bibnamefont{Lenaerts}},
  \bibinfo{journal}{Proc. Natl. Acad. Sci. USA} \textbf{\bibinfo{volume}{103}},
  \bibinfo{pages}{3490} (\bibinfo{year}{2006}).

\bibitem[{\citenamefont{Ohtsuki et~al.}(2006)\citenamefont{Ohtsuki, Hauert,
  Lieberman, and Nowak}}]{ohtsuki_n06}
\bibinfo{author}{\bibfnamefont{H.}~\bibnamefont{Ohtsuki}},
  \bibinfo{author}{\bibfnamefont{C.}~\bibnamefont{Hauert}},
  \bibinfo{author}{\bibfnamefont{E.}~\bibnamefont{Lieberman}},
  \bibnamefont{and} \bibinfo{author}{\bibfnamefont{M.~A.} \bibnamefont{Nowak}},
  \bibinfo{journal}{Nature} \textbf{\bibinfo{volume}{441}},
  \bibinfo{pages}{502} (\bibinfo{year}{2006}).

\bibitem[{\citenamefont{G{\'o}mez-Garde{\~n}es
  et~al.}(2007)\citenamefont{G{\'o}mez-Garde{\~n}es, Campillo, Flor{\' \i}a,
  and Moreno}}]{gomez-gardenes_prl07}
\bibinfo{author}{\bibfnamefont{J.}~\bibnamefont{G{\'o}mez-Garde{\~n}es}},
  \bibinfo{author}{\bibfnamefont{M.}~\bibnamefont{Campillo}},
  \bibinfo{author}{\bibfnamefont{L.~M.} \bibnamefont{Flor{\' \i}a}},
  \bibnamefont{and} \bibinfo{author}{\bibfnamefont{Y.}~\bibnamefont{Moreno}},
  \bibinfo{journal}{Phys. Rev. Lett.} \textbf{\bibinfo{volume}{98}},
  \bibinfo{pages}{108103} (\bibinfo{year}{2007}).

\bibitem[{\citenamefont{Ohtsuki et~al.}(2007)\citenamefont{Ohtsuki, Nowak, and
  Pacheco}}]{ohtsuki_prl07}
\bibinfo{author}{\bibfnamefont{H.}~\bibnamefont{Ohtsuki}},
  \bibinfo{author}{\bibfnamefont{M.~A.} \bibnamefont{Nowak}}, \bibnamefont{and}
  \bibinfo{author}{\bibfnamefont{J.~M.} \bibnamefont{Pacheco}},
  \bibinfo{journal}{Phys. Rev. Lett.} \textbf{\bibinfo{volume}{98}},
  \bibinfo{pages}{108106} (\bibinfo{year}{2007}).

\bibitem[{\citenamefont{Fu et~al.}(2009)\citenamefont{Fu, Wu, and
  Wang}}]{fu_pre09}
\bibinfo{author}{\bibfnamefont{F.}~\bibnamefont{Fu}},
  \bibinfo{author}{\bibfnamefont{T.}~\bibnamefont{Wu}}, \bibnamefont{and}
  \bibinfo{author}{\bibfnamefont{L.}~\bibnamefont{Wang}},
  \bibinfo{journal}{Phys. Rev. E} \textbf{\bibinfo{volume}{79}},
  \bibinfo{pages}{036101} (\bibinfo{year}{2009}).

\bibitem[{\citenamefont{Pusch et~al.}(2008)\citenamefont{Pusch, Weber, and
  Porto}}]{pusch_pre08}
\bibinfo{author}{\bibfnamefont{A.}~\bibnamefont{Pusch}},
  \bibinfo{author}{\bibfnamefont{S.}~\bibnamefont{Weber}}, \bibnamefont{and}
  \bibinfo{author}{\bibfnamefont{M.}~\bibnamefont{Porto}},
  \bibinfo{journal}{Phys. Rev. E} \textbf{\bibinfo{volume}{77}},
  \bibinfo{pages}{036120} (\bibinfo{year}{2008}).

\bibitem[{\citenamefont{Traulsen et~al.}(2009)\citenamefont{Traulsen, Hauert,
  H., Nowak, and Sigmund}}]{traulsen_pnas09}
\bibinfo{author}{\bibfnamefont{A.}~\bibnamefont{Traulsen}},
  \bibinfo{author}{\bibfnamefont{C.}~\bibnamefont{Hauert}},
  \bibinfo{author}{\bibfnamefont{D.~S.} \bibnamefont{H.}},
  \bibinfo{author}{\bibfnamefont{M.~A.} \bibnamefont{Nowak}}, \bibnamefont{and}
  \bibinfo{author}{\bibfnamefont{K.}~\bibnamefont{Sigmund}},
  \bibinfo{journal}{Proc. Natl. Acad. Sci. USA} \textbf{\bibinfo{volume}{106}},
  \bibinfo{pages}{709} (\bibinfo{year}{2009}).

\bibitem[{\citenamefont{Van~Segbroeck et~al.}(2009)\citenamefont{Van~Segbroeck,
  Santos, Lenaerts, and Pacheco}}]{van-segbroeck_prl09}
\bibinfo{author}{\bibfnamefont{S.}~\bibnamefont{Van~Segbroeck}},
  \bibinfo{author}{\bibfnamefont{F.~C.} \bibnamefont{Santos}},
  \bibinfo{author}{\bibfnamefont{T.}~\bibnamefont{Lenaerts}}, \bibnamefont{and}
  \bibinfo{author}{\bibfnamefont{J.~M.} \bibnamefont{Pacheco}},
  \bibinfo{journal}{Phys. Rev. Lett.} \textbf{\bibinfo{volume}{102}},
  \bibinfo{pages}{058105} (\bibinfo{year}{2009}).

\bibitem[{\citenamefont{Lee et~al.}(2011)\citenamefont{Lee, Holme, and
  Wu}}]{lee_s_prl11}
\bibinfo{author}{\bibfnamefont{S.}~\bibnamefont{Lee}},
  \bibinfo{author}{\bibfnamefont{P.}~\bibnamefont{Holme}}, \bibnamefont{and}
  \bibinfo{author}{\bibfnamefont{Z.-X.} \bibnamefont{Wu}},
  \bibinfo{journal}{Phys. Rev. Lett.} \textbf{\bibinfo{volume}{106}},
  \bibinfo{pages}{028702} (\bibinfo{year}{2011}).

\bibitem[{\citenamefont{Poncela et~al.}(2011)\citenamefont{Poncela,
  G{\'o}mez-Garde{\~n}es, and Moreno}}]{poncela_pre11}
\bibinfo{author}{\bibfnamefont{J.}~\bibnamefont{Poncela}},
  \bibinfo{author}{\bibfnamefont{J.}~\bibnamefont{G{\'o}mez-Garde{\~n}es}},
  \bibnamefont{and} \bibinfo{author}{\bibfnamefont{Y.}~\bibnamefont{Moreno}},
  \bibinfo{journal}{Phys. Rev. E} \textbf{\bibinfo{volume}{83}},
  \bibinfo{pages}{057101} (\bibinfo{year}{2011}).

\bibitem[{\citenamefont{Buesser and Tomassini}(2012)}]{buesser_pre12}
\bibinfo{author}{\bibfnamefont{P.}~\bibnamefont{Buesser}} \bibnamefont{and}
  \bibinfo{author}{\bibfnamefont{M.}~\bibnamefont{Tomassini}},
  \bibinfo{journal}{Phys. Rev. E} \textbf{\bibinfo{volume}{85}},
  \bibinfo{pages}{016107} (\bibinfo{year}{2012}).

\bibitem[{\citenamefont{Yang et~al.}(2012)\citenamefont{Yang, Wu, and
  Du}}]{yang_hx_epl12}
\bibinfo{author}{\bibfnamefont{H.-X.} \bibnamefont{Yang}},
  \bibinfo{author}{\bibfnamefont{Z.-X.} \bibnamefont{Wu}}, \bibnamefont{and}
  \bibinfo{author}{\bibfnamefont{W.-B.} \bibnamefont{Du}},
  \bibinfo{journal}{EPL} \textbf{\bibinfo{volume}{99}}, \bibinfo{pages}{10006}
  (\bibinfo{year}{2012}).

\bibitem[{\citenamefont{Chen and Lai}(2012)}]{chen_yz_pre12}
\bibinfo{author}{\bibfnamefont{Y.-Z.} \bibnamefont{Chen}} \bibnamefont{and}
  \bibinfo{author}{\bibfnamefont{Y.-C.} \bibnamefont{Lai}},
  \bibinfo{journal}{Phys. Rev. E} \textbf{\bibinfo{volume}{86}},
  \bibinfo{pages}{045101(R)} (\bibinfo{year}{2012}).

\bibitem[{\citenamefont{Pinheiro et~al.}(2012)\citenamefont{Pinheiro, Santos,
  and Pacheco}}]{pinheiro_njp12}
\bibinfo{author}{\bibfnamefont{F.}~\bibnamefont{Pinheiro}},
  \bibinfo{author}{\bibfnamefont{F.}~\bibnamefont{Santos}}, \bibnamefont{and}
  \bibinfo{author}{\bibfnamefont{J.}~\bibnamefont{Pacheco}},
  \bibinfo{journal}{New J. Phys.} \textbf{\bibinfo{volume}{14}},
  \bibinfo{pages}{073035} (\bibinfo{year}{2012}).

\bibitem[{\citenamefont{Allen et~al.}(2012)\citenamefont{Allen, Traulsen,
  Tarnita, and Nowak}}]{allen_jtb12}
\bibinfo{author}{\bibfnamefont{B.}~\bibnamefont{Allen}},
  \bibinfo{author}{\bibfnamefont{A.}~\bibnamefont{Traulsen}},
  \bibinfo{author}{\bibfnamefont{C.~E.} \bibnamefont{Tarnita}},
  \bibnamefont{and} \bibinfo{author}{\bibfnamefont{M.~A.} \bibnamefont{Nowak}},
  \bibinfo{journal}{J. Theor. Biol.} \textbf{\bibinfo{volume}{299}},
  \bibinfo{pages}{97} (\bibinfo{year}{2012}).

\bibitem[{\citenamefont{Kuperman and Risau-Gusman}(2012)}]{kuperman_pre12}
\bibinfo{author}{\bibfnamefont{M.~N.} \bibnamefont{Kuperman}} \bibnamefont{and}
  \bibinfo{author}{\bibfnamefont{S.}~\bibnamefont{Risau-Gusman}},
  \bibinfo{journal}{Phys. Rev. E} \textbf{\bibinfo{volume}{86}},
  \bibinfo{pages}{016104} (\bibinfo{year}{2012}).

\bibitem[{\citenamefont{Allen and Nowak}(2012)}]{allen_jtb12b}
\bibinfo{author}{\bibfnamefont{B.}~\bibnamefont{Allen}} \bibnamefont{and}
  \bibinfo{author}{\bibfnamefont{M.~A.} \bibnamefont{Nowak}},
  \bibinfo{journal}{J. Theor. Biol.} \textbf{\bibinfo{volume}{311}},
  \bibinfo{pages}{28} (\bibinfo{year}{2012}).

\bibitem[{\citenamefont{Li et~al.}(2012)\citenamefont{Li, Iqbal, Chen, and
  Abbott}}]{li_q_njp12}
\bibinfo{author}{\bibfnamefont{Q.}~\bibnamefont{Li}},
  \bibinfo{author}{\bibfnamefont{A.}~\bibnamefont{Iqbal}},
  \bibinfo{author}{\bibfnamefont{M.}~\bibnamefont{Chen}}, \bibnamefont{and}
  \bibinfo{author}{\bibfnamefont{D.}~\bibnamefont{Abbott}},
  \bibinfo{journal}{New J. Phys.} \textbf{\bibinfo{volume}{14}},
  \bibinfo{pages}{103034} (\bibinfo{year}{2012}).

\bibitem[{\citenamefont{Tanimoto et~al.}(2012)\citenamefont{Tanimoto, Brede,
  and Yamauchi}}]{tanimoto_pre12}
\bibinfo{author}{\bibfnamefont{J.}~\bibnamefont{Tanimoto}},
  \bibinfo{author}{\bibfnamefont{M.}~\bibnamefont{Brede}}, \bibnamefont{and}
  \bibinfo{author}{\bibfnamefont{A.}~\bibnamefont{Yamauchi}},
  \bibinfo{journal}{Phys. Rev. E} \textbf{\bibinfo{volume}{85}},
  \bibinfo{pages}{032101} (\bibinfo{year}{2012}).

\bibitem[{\citenamefont{Fu and Nowak}(2012)}]{fu_jsp12}
\bibinfo{author}{\bibfnamefont{F.}~\bibnamefont{Fu}} \bibnamefont{and}
  \bibinfo{author}{\bibfnamefont{M.}~\bibnamefont{Nowak}}, \bibinfo{journal}{J.
  Stat. Phys.}{in press}  (\bibinfo{year}{2013}).

\bibitem[{\citenamefont{Portillo}(2012)}]{portillo_pre12}
\bibinfo{author}{\bibfnamefont{I.~G.} \bibnamefont{Portillo}},
  \bibinfo{journal}{Phys. Rev. E} \textbf{\bibinfo{volume}{86}},
  \bibinfo{pages}{051108} (\bibinfo{year}{2012}).

\bibitem[{\citenamefont{Szolnoki et~al.}(2011)\citenamefont{Szolnoki, Xie,
  Wang, and Perc}}]{szolnoki_epl11}
\bibinfo{author}{\bibfnamefont{A.}~\bibnamefont{Szolnoki}},
  \bibinfo{author}{\bibfnamefont{N.-G.} \bibnamefont{Xie}},
  \bibinfo{author}{\bibfnamefont{C.}~\bibnamefont{Wang}}, \bibnamefont{and}
  \bibinfo{author}{\bibfnamefont{M.}~\bibnamefont{Perc}},
  \bibinfo{journal}{EPL} \textbf{\bibinfo{volume}{96}}, \bibinfo{pages}{38002}
  (\bibinfo{year}{2011}).

\bibitem[{\citenamefont{Hill et~al.}(2010)\citenamefont{Hill, Rand,
  Nowak, and Christakis}}]{hill_prsb10}
\bibinfo{author}{\bibfnamefont{A.~L.}~\bibnamefont{Hill}},
  \bibinfo{author}{\bibfnamefont{D.~G.} \bibnamefont{Rand}},
  \bibinfo{author}{\bibfnamefont{M.~A.}~\bibnamefont{Nowak}}, \bibnamefont{and}
  \bibinfo{author}{\bibfnamefont{N.~A.}~\bibnamefont{Christakis}},
  \bibinfo{journal}{Proc. R. Soc. B} \textbf{\bibinfo{volume}{277}}, \bibinfo{pages}{3827}
  (\bibinfo{year}{2010}).

\bibitem[{\citenamefont{Rand et~al.}(2012)\citenamefont{Rand,
  Greene, and Nowak}}]{rand_n12}
\bibinfo{author}{\bibfnamefont{D.~G.} \bibnamefont{Rand}},
  \bibinfo{author}{\bibfnamefont{J.~D.}~\bibnamefont{Greene}}, \bibnamefont{and}
  \bibinfo{author}{\bibfnamefont{M.~A.}~\bibnamefont{Nowak}},
  \bibinfo{journal}{Nature} \textbf{\bibinfo{volume}{489}}, \bibinfo{pages}{427}
  (\bibinfo{year}{2012}).

\bibitem[{\citenamefont{Brede}(2011)}]{brede_epl11}
\bibinfo{author}{\bibfnamefont{M.}~\bibnamefont{Brede}}, \bibinfo{journal}{EPL}
  \textbf{\bibinfo{volume}{94}}, \bibinfo{pages}{30003} (\bibinfo{year}{2011}).

\bibitem[{\citenamefont{Vukov et~al.}(2012)\citenamefont{Vukov, Santos, and
  Pacheco}}]{vukov_njp12}
\bibinfo{author}{\bibfnamefont{J.}~\bibnamefont{Vukov}},
  \bibinfo{author}{\bibfnamefont{F.}~\bibnamefont{Santos}}, \bibnamefont{and}
  \bibinfo{author}{\bibfnamefont{J.}~\bibnamefont{Pacheco}},
  \bibinfo{journal}{New J. Phys.} \textbf{\bibinfo{volume}{14}},
  \bibinfo{pages}{063031} (\bibinfo{year}{2012}).

\bibitem[{\citenamefont{Brede}(2012)}]{brede_acs12}
\bibinfo{author}{\bibfnamefont{M.}~\bibnamefont{Brede}},
  \bibinfo{journal}{Advs. Complex Syst.} \textbf{\bibinfo{volume}{15}},
  \bibinfo{pages}{1250074} (\bibinfo{year}{2012}).

\bibitem[{\citenamefont{Szolnoki and Perc}(2012)}]{szolnoki_pre12}
\bibinfo{author}{\bibfnamefont{A.}~\bibnamefont{Szolnoki}} \bibnamefont{and}
  \bibinfo{author}{\bibfnamefont{M.}~\bibnamefont{Perc}},
  \bibinfo{journal}{Phys. Rev. E} \textbf{\bibinfo{volume}{85}},
  \bibinfo{pages}{026104} (\bibinfo{year}{2012}).

\bibitem[{\citenamefont{Nowak and Sigmund}(1992)}]{nowak_n92a}
\bibinfo{author}{\bibfnamefont{M.~A.} \bibnamefont{Nowak}} \bibnamefont{and}
  \bibinfo{author}{\bibfnamefont{K.}~\bibnamefont{Sigmund}},
  \bibinfo{journal}{Nature} \textbf{\bibinfo{volume}{355}},
  \bibinfo{pages}{250} (\bibinfo{year}{1992}).

\bibitem[{\citenamefont{Wedekind and Milinski}(1996)}]{wedekind_pnas96}
\bibinfo{author}{\bibfnamefont{C.}~\bibnamefont{Wedekind}} \bibnamefont{and}
  \bibinfo{author}{\bibfnamefont{M.}~\bibnamefont{Milinski}},
  \bibinfo{journal}{Proc. Natl. Acad. Sci. USA} \textbf{\bibinfo{volume}{93}},
  \bibinfo{pages}{2686} (\bibinfo{year}{1996}).

\bibitem[{\citenamefont{Axelrod and Hamilton}(1981)}]{axelrod_s81}
\bibinfo{author}{\bibfnamefont{R.}~\bibnamefont{Axelrod}} \bibnamefont{and}
  \bibinfo{author}{\bibfnamefont{W.~D.} \bibnamefont{Hamilton}},
  \bibinfo{journal}{Science} \textbf{\bibinfo{volume}{211}},
  \bibinfo{pages}{1390} (\bibinfo{year}{1981}).

\bibitem[{\citenamefont{Gracia-L{\'a}zaro
  et~al.}(2012)\citenamefont{Gracia-L{\'a}zaro, Cuesta, S{\'a}nchez, and
  Moreno}}]{gracia-lazaro_srep12}
\bibinfo{author}{\bibfnamefont{C.}~\bibnamefont{Gracia-L{\'a}zaro}},
  \bibinfo{author}{\bibfnamefont{J.}~\bibnamefont{Cuesta}},
  \bibinfo{author}{\bibfnamefont{A.}~\bibnamefont{S{\'a}nchez}},
  \bibnamefont{and} \bibinfo{author}{\bibfnamefont{Y.}~\bibnamefont{Moreno}},
  \bibinfo{journal}{Sci. Rep.} \textbf{\bibinfo{volume}{2}},
  \bibinfo{pages}{325} (\bibinfo{year}{2012}).

\bibitem[{\citenamefont{Szab{\'o} et~al.}(2004)\citenamefont{Szab{\'o},
  Szolnoki, and Izs{\'a}k}}]{szabo_jpa04}
\bibinfo{author}{\bibfnamefont{G.}~\bibnamefont{Szab{\'o}}},
  \bibinfo{author}{\bibfnamefont{A.}~\bibnamefont{Szolnoki}}, \bibnamefont{and}
  \bibinfo{author}{\bibfnamefont{R.}~\bibnamefont{Izs{\'a}k}},
  \bibinfo{journal}{J. Phys. A: Math. Gen.} \textbf{\bibinfo{volume}{37}},
  \bibinfo{pages}{2599} (\bibinfo{year}{2004}).

\bibitem[{\citenamefont{Szab{\'o} et~al.}(2005)\citenamefont{Szab{\'o}, Vukov,
  and Szolnoki}}]{szabo_pre05}
\bibinfo{author}{\bibfnamefont{G.}~\bibnamefont{Szab{\'o}}},
  \bibinfo{author}{\bibfnamefont{J.}~\bibnamefont{Vukov}}, \bibnamefont{and}
  \bibinfo{author}{\bibfnamefont{A.}~\bibnamefont{Szolnoki}},
  \bibinfo{journal}{Phys. Rev. E} \textbf{\bibinfo{volume}{72}},
  \bibinfo{pages}{047107} (\bibinfo{year}{2005}).

\bibitem[{\citenamefont{Vukov et~al.}(2006)\citenamefont{Vukov, Szab{\'o}, and
  Szolnoki}}]{vukov_pre06}
\bibinfo{author}{\bibfnamefont{J.}~\bibnamefont{Vukov}},
  \bibinfo{author}{\bibfnamefont{G.}~\bibnamefont{Szab{\'o}}},
  \bibnamefont{and} \bibinfo{author}{\bibfnamefont{A.}~\bibnamefont{Szolnoki}},
  \bibinfo{journal}{Phys. Rev. E} \textbf{\bibinfo{volume}{73}},
  \bibinfo{pages}{067103} (\bibinfo{year}{2006}).

\bibitem[{\citenamefont{Vilone et~al.}(2011)\citenamefont{Vilone, S{\'a}nchez,
  and G{\'o}mez-Garde{\~n}es}}]{vilone_jsm11}
\bibinfo{author}{\bibfnamefont{D.}~\bibnamefont{Vilone}},
  \bibinfo{author}{\bibfnamefont{A.}~\bibnamefont{S{\'a}nchez}},
  \bibnamefont{and}
  \bibinfo{author}{\bibfnamefont{J.}~\bibnamefont{G{\'o}mez-Garde{\~n}es}},
  \bibinfo{journal}{J. Stat. Mech} \textbf{\bibinfo{volume}{2011}},
  \bibinfo{pages}{P04019} (\bibinfo{year}{2011}).

\bibitem[{\citenamefont{Barab{\'a}si and Albert}(1999)}]{barabasi_s99}
\bibinfo{author}{\bibfnamefont{A.-L.} \bibnamefont{Barab{\'a}si}}
  \bibnamefont{and} \bibinfo{author}{\bibfnamefont{R.}~\bibnamefont{Albert}},
  \bibinfo{journal}{Science} \textbf{\bibinfo{volume}{286}},
  \bibinfo{pages}{509} (\bibinfo{year}{1999}).

\bibitem[{\citenamefont{Santos and Pacheco}(2006)}]{santos_jeb06}
\bibinfo{author}{\bibfnamefont{F.~C.} \bibnamefont{Santos}} \bibnamefont{and}
  \bibinfo{author}{\bibfnamefont{J.~M.} \bibnamefont{Pacheco}},
  \bibinfo{journal}{J. Evol. Biol.} \textbf{\bibinfo{volume}{19}},
  \bibinfo{pages}{726} (\bibinfo{year}{2006}).

\bibitem[{\citenamefont{Tomassini et~al.}(2007)\citenamefont{Tomassini, Luthi,
  and Pestelacci}}]{tomassini_ijmpc07}
\bibinfo{author}{\bibfnamefont{M.}~\bibnamefont{Tomassini}},
  \bibinfo{author}{\bibfnamefont{L.}~\bibnamefont{Luthi}}, \bibnamefont{and}
  \bibinfo{author}{\bibfnamefont{E.}~\bibnamefont{Pestelacci}},
  \bibinfo{journal}{Int. J. Mod. Phys. C} \textbf{\bibinfo{volume}{18}},
  \bibinfo{pages}{1173} (\bibinfo{year}{2007}).

\bibitem[{\citenamefont{Wu et~al.}(2007)\citenamefont{Wu, Guan, Xu, and
  Y.-H.Wang}}]{wu_zx_pa07}
\bibinfo{author}{\bibfnamefont{Z.-X.} \bibnamefont{Wu}},
  \bibinfo{author}{\bibfnamefont{J.-Y.} \bibnamefont{Guan}},
  \bibinfo{author}{\bibfnamefont{X.-J.} \bibnamefont{Xu}}, \bibnamefont{and}
  \bibinfo{author}{\bibnamefont{Y.-H.Wang}}, \bibinfo{journal}{Physica A}
  \textbf{\bibinfo{volume}{379}}, \bibinfo{pages}{672} (\bibinfo{year}{2007}).

\bibitem[{\citenamefont{Masuda}(2007)}]{masuda_prsb07}
\bibinfo{author}{\bibfnamefont{N.}~\bibnamefont{Masuda}},
  \bibinfo{journal}{Proc. R. Soc. B} \textbf{\bibinfo{volume}{274}},
  \bibinfo{pages}{1815} (\bibinfo{year}{2007}).

\bibitem[{\citenamefont{Szolnoki et~al.}(2008)\citenamefont{Szolnoki, Perc, and
  Danku}}]{szolnoki_pa08}
\bibinfo{author}{\bibfnamefont{A.}~\bibnamefont{Szolnoki}},
  \bibinfo{author}{\bibfnamefont{M.}~\bibnamefont{Perc}}, \bibnamefont{and}
  \bibinfo{author}{\bibfnamefont{Z.}~\bibnamefont{Danku}},
  \bibinfo{journal}{Physica A} \textbf{\bibinfo{volume}{387}},
  \bibinfo{pages}{2075} (\bibinfo{year}{2008}).

\bibitem[{\citenamefont{Davidson and Begley}(2012)}]{davidson_12}
\bibinfo{author}{\bibfnamefont{R.~J.}~\bibnamefont{Davidson}} \bibnamefont{and}
  \bibinfo{author}{\bibfnamefont{S.}~\bibnamefont{Begley}},
  \emph{\bibinfo{title}{The emotional life of your brain}}
  (\bibinfo{publisher}{Hudson Street Press},
  \bibinfo{address}{New York, U.S.}, \bibinfo{year}{2012}).

\end{thebibliography}
\end{document}